\documentclass[12pt]{article}
\pdfoutput=1
\usepackage{amsbsy}
\usepackage{amsfonts}
\usepackage{amsmath,amssymb}
\usepackage{bbold}
\usepackage{pdfpages}

\newcommand{\sect}[1]{\setcounter{equation}{0}\section{#1}}

\newcommand{\eq}{\begin{equation}}
\newcommand{\eqa}{\begin{eqnarray}}  
\newcommand{\en}{\end{equation}}
\newcommand{\ena}{\end{eqnarray}}
\newcommand{\enn}{\nonumber \end{equation}}

\def\sk{\vskip .4cm}
\def\noi{\noindent}
\def\om{\omega}
\def\al{\alpha}
\def\be{\beta}
\def\ga{\gamma}
\def\Ga{\Gamma}

\let \si\sigma
\let \part\partial

\def\unquarto{{1 \over 4}}

\def\unmezzo{{1 \over 2}}
\def\epsi{\varepsilon}
\def\we{\wedge}

\def\de{\delta}

\def\Gt{{\tilde G}}

\def\part{\partial}

\def\sk{\vskip .4cm}

\def\noi{\noindent}

\def\pdyN{{\partial \over \partial y^N}}

\def\X0{X^0}

\def\om{\omega}
\def\omo{{\buildrel \circ \over \omega}}
\def\al{\alpha}
\def\ga{\gamma}

\def\unquarto{{1 \over 4}}
\def\unmezzo{{1 \over 2}}
\def\epsi{\varepsilon}
\def\epsibold{{\bf \epsilon}}

\def\psib{{\bar \psi}}

\def\we{\wedge}

\def\de{\delta}
\def\CABC{{C^A}_{BC}}
\def\CCAB{{C^C}_{AB}}

\def\Gt{{\tilde G}}

\def\La{\Lambda}

\def\Dcal{{\cal D}}
\def\Rcal{{\cal R}}

\def\square{{\,\lower0.9pt\vbox{\hrule \hbox{\vrule height 0.2 cm
\hskip 0.2 cm \vrule height 0.2 cm}\hrule}\,}}

\def\epsilonbar{{\bar \epsilon}}
\def\thetabar{{\bar \theta}}

\def\Ftilde{\tilde F}
\def\Gtilde{\tilde G}

\def\psibar{\bar \psi}
\def\epsilonbar{\bar \epsilon}
\def\chibar{\bar \chi}
\def\rhobar{\bar \rho}

\def\Om{\Omega}
\def\Qbar{\bar Q}

\def\Sigmabar{\overline \Sigma}

\def\Rbold{{\bf R}}

\def\Ombold{{\bf \Om}}
\def\onebold{{\bf 1}}
\def\epsibold{\boldsymbol {\epsilon}}

\def\Gammabold{{\bf \Gamma}}

\def\Gbold{{\bf G}}

\def\One{\mathbb{1}}

\def\dright{\stackrel{\rightarrow}{\part}}
\def\dleft{\stackrel{\leftarrow}{\part}}

\begin{document}

\begin{titlepage}
\rightline{ARC-18-02}
\vskip 2em
\begin{center}
{\Large \bf Supergravity in the group-geometric framework: a primer  } \\[3em]

\vskip 0.5cm

{\bf
Leonardo Castellani}
\medskip

\vskip 0.5cm

{\sl Dipartimento di Scienze e Innovazione Tecnologica
\\Universit\`a del Piemonte Orientale, viale T. Michel 11, 15121 Alessandria, Italy\\ [.5em] INFN, Sezione di 
Torino, via P. Giuria 1, 10125 Torino, Italy\\ [.5em]
Arnold-Regge Center, via P. Giuria 1, 10125 Torino, Italy
}\\ [4em]
\end{center}

\begin{abstract}
\sk

We review the group-geometric approach to supergravity theories, in the perspective of
recent developments and applications. Usual diffeomorphisms, gauge symmetries and supersymmetries are unified as superdiffeomorphisms in a supergroup manifold.  Integration on supermanifolds is briefly revisited, and used as a tool to provide a bridge between component and superspace actions. As an illustration of the constructive techniques, the cases of $d=3,4$ off-shell supergravities and $d=5$ Chern-Simons supergravity are discussed in detail. A cursory account of $d=10+2$ supergravity is also included. We recall a covariant canonical formalism, well adapted to theories described by Lagrangians $d$-forms, that allows to define a form hamiltonian and to recast constrained hamiltonian systems in a covariant form language. 
Finally, group geometry and properties of spinors and gamma matrices in $d=s+t$ dimensions are summarized in Appendices.

\end{abstract}

\vskip 1cm\sk\sk
\noi {\it {\small Based on lectures given at : the Graduate School of the Physics Dept , W\"urzburg, 2011; 
Corfu Workshop 2013; LACES 2013, Lezioni Avanzate di Campi e Stringhe, GGI, Florence;
CERN-SEENET PhD Training Program, Belgrade, 2015; GGI 2016 Workshop "Supergravity, what next?", Florence. }}
\sk\sk
 \noi \hrule \vskip .2cm \noi {\small
leonardo.castellani@uniupo.it}

\end{titlepage}

\newpage
\setcounter{page}{1}

\tableofcontents

\sect{Introduction}

Field theories with local symmetries are the theoretical tool ``par excellence" to describe elementary particles and their interactions. Fields depend on spacetime coordinates $x$ and have specific transformation properties under local symmetries. For example the fields of the standard model  transform under gauge symmetries as gauge or matter fields, and
in a way dictated by the group representation they belong to. The fields in gravity theories also transform under diffeomorphisms according to their tensorial character. The essential difference between these two types of local transformations, in their infinitesimal versions, is that diffeomorphisms always contain a derivative of the field,  which is absent in gauge transformations. This is simply due to the fact that general coordinate transformations relate fields at different spacetime points, whereas gauge transformations relate fields at the same spacetime point.
\sk
On the other hand, the idea of unifying gravity with gauge theories starting from their symmetry structure has an old and well-motivated history, culminating in the relatively recent AdS/CFT or
gauge/gravity correspondences. 

Even on the classical level we can provide a unified description
of diffeomorphisms and gauge transformations. For this we need  a group geometrical framework.

To set the stage we consider as basic fields of the theory the components of the vielbein 
 one-form $\sigma^A=\sigma(z)^A_{~\Lambda} dz^\Lambda$ on the manifold of a Lie group $G$,  {\small {\it A}} being an index in the $G$ Lie algebra, and $z^\Lambda$ the coordinates of the group manifold.
 This vielbein satisfies the Cartan-Maurer (CM) equations\footnote{a short summary of group manifold geometry is given in Appendix A.}
\eq
d \sigma^A + {1 \over 2} C^A_{BC} ~\sigma^B \wedge \sigma^C =0 \label{CM}
\en
where $C^A_{BC}$ are the structure constants of the $G$ Lie algebra.  
The $G$ vielbein $\sigma^A (z)$  has a fixed dependence on the coordinates $z$, and cannot
therefore play the r\^ole of a dynamical object. We must consider then a ``soft" group manifold, 
diffeomorphic to $G$ and denoted by ${\tilde G}$,
with a vielbein $\sigma^A$ not satisfying anymore the CM equations. The amount of deformation from the original ``rigid" group manifold is measured by the curvature two-form:
\eq
R^A \equiv d \sigma^A + {1 \over 2} C^A_{BC} ~\sigma^B \wedge \sigma^C  \label{Gcurvature}
\en
Thus the soft $\Gtilde$ can fluctuate around the rigid $G$ manifold (with $R^A=0$), in the same way spacetime of general relativity can fluctuate around flat Minkowski spacetime. Tangent vectors
on $\Gtilde$, dual to the vielbein $\sigma^A$, are denoted by $t_B$, so that $\sigma^A (t_B)=\delta^A_B$.
\sk
Diffeomorphisms along tangent vectors $\epsi=\epsi^A t_A$ on $\Gtilde$ are generated by the Lie derivative $\ell_\epsi$.  When applied to the $\Gtilde$ vielbein, the variation under diffeomorphisms takes the suggestive form:
\eq
\ell_\epsi \sigma^A = d \epsi^A + C^A_{BC} \sigma^B \epsi^C + \iota_\epsi R^A \label{Lieder}
\en
(where $\iota_\epsi$ is the contraction operator, see Appendix A) and one recognizes on the right-hand side the $G$-covariant derivative of the infinitesimal parameter 
 $\epsi^A$ plus a curvature term. When the curvature term vanishes, i.e. when 
 $\iota_\epsi R^A=0$, the diffeomorphism
 takes the form of a {\it gauge transformation},  and the
 curvature is said to be {\it horizontal}
 along the $t_A$'s entering the sum in $\epsi=\epsi^A t_A$.
 
 Thus in group manifold geometry {\it gauge transformations} can be interpreted as {\it particular
 diffeomorphisms}, along the directions on which the curvatures are horizontal.
 
 To make the exposition more pedagogical, the group manifold approach will be developed within the basic examples of gravity and supergravity in $d=4$.
 
 The paper is organized as follows. Sections 2 and 3 deal with (first-order) gravity and
 supergravity in $d=4$, and serve as an introduction to the group-geometric framework.
 The original references, where this approach was first proposed, 
 are given in \cite{gm11}-\cite{gm14}. Reviews can be found 
 in \cite{gm21}-\cite{gm23}. In Section 4 we recall basic results in supermanifold integration
 (see for ex. \cite{w1} for a recent review, or \cite{voronov} for a textbook),
 and new developments concerning integral forms, discussed in ref.s  \cite{if1}-\cite{if5}.
 Section 5 summarizes the building rules of $d$-form Lagrangians, applied in
 subsequent examples. Section 6 extends the formalism to $p$-forms fields by
 considering Free Differential Algebras (FDA), first introduced in the context of
 supergravity theories in \cite{DFd11}. Off-shell $d=3$ supergravity is recast in the
 group manifold setting in Section 7 and used in Section 8 to
 establish a bridge between the component and superspace actions, following \cite{if3}.
 Off-shell (new minimal) $d=4$ supergravity in the group manifold setting is discussed in Section 9, 
 based on ref. \cite{DFTvN}. In Section 10 we provide selected examples of gauge supergravities,
 in odd dimensions (Chern-Simons supergravities, for a review see for example \cite{Zanelli}) and
 in even dimensions (generalizations of the Mac Dowell-Mansouri action \cite{MDM}). In these
 theories supersymmetry ``lives on the fiber",  i.e. is part of a {\it gauged superalgebra}, and is
 not interpreted as a superdiffeomorphism. Finally, Section 11 recalls a covariant
 hamiltonian formalism well adapted to $d$-form Lagrangians \cite{CCF1}-\cite{CCF5}, with an
 application to pure vierbein gravity first discussed in \cite{CCF1}. The Appendices contain a minireview on group geometry, and properties of spinors and gamma matrices in $d=s+t$ dimensions.

 \sect{The first example: Poincar\'e gravity}
 
 \subsection{Soft Poincar\'e manifold} 
 
Gravity in first order vierbein formalism can be recast in a group geometric setting as follows.
Consider $G$ = Poincar\'e group, and denote the vielbein on the $\Gtilde$ manifold as
$\sigma^A = (V^a,\omega^{ab})$. The index $A=(a,ab)$ runs on the translations and Lorentz rotations of the Poincar\'e Lie algebra:
 \eqa
 & &  [P_a,P_b] =0  \label{PoincarePP} \\
 & & [M_{ab},M_{cd}]= -{1 \over 2} (\eta_{ad} M_{bc} + \eta_{bc} M_{ad} -\eta_{ac} M_{bd} -\eta_{bd} M_{ac}) \label{PoincareMM} \\
 & & [M_{ab},P_c]= -{1 \over 2} ( \eta_{bc} P_{a} - \eta_{ac} P_{b}) \label{PoincareMP}
  \ena 
  \noi $\eta$ being the flat Minkowski metric. The $V^a$ and $\omega^{ab}$ components of the $\Gtilde$ vielbein are identified
  with the vierbein and the spin connection. The curvature
  two-form defined as in (\ref{Gcurvature}) becomes
  \eqa
 & & R^a = dV^a - \omega^{ab} \wedge V^c \eta_{bc} \label{defRa}\\
 & & R^{ab} = d \omega^{ab} - \omega^{ac} \wedge \omega^{bd} \eta_{cd} \label{defRab}
 \ena
and one recognizes the familiar expressions for the torsion and the Lorentz curvature.
Taking the exterior derivative of these definitions yields the Bianchi identities (BI):
\eqa
& & dR^a = - R^{ab} \we V^c \eta_{bc} + \omega^{ab} \we R^c \eta_{bc}  ~~\longrightarrow 
 \Dcal R^a = - R^{ab} \we V^c \eta_{bc}    \label{bianchiRa}\\
& & dR^{ab} = (-R^{ac} \we \omega^{bd} + \omega^{ac} \we R^{bd} ) \eta_{cd} 
\longrightarrow \Dcal R^{ab} =0 \label{bianchiRab}
\ena
where $\Dcal$ is the Lorentz covariant exterior derivative.

At this stage all the fields depend on the $\Gtilde$ manifold coordinates, corresponding
to the generators of the Lie algebra: thus $V^a=V^a (x,y)$, $\omega^{ab} = \omega^{ab}(x,y)$
where the coordinates $x^a$, corresponding to the translations $P_a$, describe usual spacetime, whereas $y^{ab}$ are the coordinates in the ``Lorentz directions", corresponding to the $SO(1,3)$ rotations generated by $M_{ab}$. Moreover the one-forms $V^a$, $\omega^{ab}$ live on the whole
$\Gtilde$, and therefore can be expanded as:
 \eqa
  & & V^a = V^a_\mu (x,y) dx^\mu + V^a_{\mu\nu} (x,y) dy^{\mu\nu} \\
  & & \omega^{ab} = \omega^{ab}_\mu (x,y) dx^\mu + \omega^{ab}_{\mu\nu} (x,y)dy^{\mu\nu} 
  \ena
It would seem that we have an embarassment of riches, with unwanted extra fields 
$V^a_{\mu\nu}, \omega^{ab}_{\mu\nu}$ and dependence of all the fields on extra coordinates $y$.  

\subsection{Group manifold action}

The overabundance of field components, and their dependence on $y$ coordinates
can be tamed by defining an appropriate action principle. To end up with a geometrical theory in
four spacetime dimensions, we first construct a 4-form Lagrangian $L$ made out of the 
$\Gtilde$ vielbein $\sigma^A$ and its curvature $R^A$, according to some 
building rules to be discussed later (Section 5). The Lagrangian for Poincar\'e gravity is given by:
\eq
L = R^{ab} \wedge V^c \wedge V^d \epsilon_{abcd} \label{LPoincare}
\en
We then define an action by integrating this Lagrangian on a 4-dimensional submanifold $M^4$
of the $\Gtilde$ manifold, spanned by the $x$ coordinates. 

Integration on submanifolds $M^d$ of a $d$-form $L$ that lives on a $g$-dimensional bigger space $\Gtilde$ can be performed as follows: we multiply $L$ by the {\it Poincar\'e dual} of $M^d$, a (singular) closed  ({\it g-d})-form $\eta_{M^d}$ that localizes the Lagrangian on the submanifold $M^d$, and integrate the resulting $g$-form on the whole $\Gtilde$. Thus the group manifold action is given by
\eq
 S= \int_{\Gtilde} L \wedge  \eta_{M^d}    \label{Gintegral}
\en
The fields of the theory are those contained in $L$, i.e. the $\Gtilde$ vielbein components, and the
embedding functions that define the $M^d$ submanifold of $\Gtilde$, present in $\eta_{M^d}$. We will see
in Section 2.6 that the embedding functions do not enter the field equations obtained from the variation of (\ref{Gintegral}).

In our example the group manifold action is the integral  of a 10-form on $\Gtilde$ = soft Poincar\'e manifold:
\eq
S=  \int_{\Gtilde} R^{ab} \wedge V^c \wedge V^d \epsilon_{abcd} \wedge  \eta_{M^4}    \label{GintegralP}
\en

\subsection{Spacetime action}

Consider now the action (\ref{GintegralP}), but with a {\it particular choice} of $\eta$ given by  the 6-form
 \eq
  \eta_{M^4} =  \delta (y^{12}) \delta (y^{13}) \cdots \delta (y^{34}) dy^{12} \wedge dy^{13} \wedge \cdots \wedge dy^{34}  \label{eta6form}
  \en
 Integration on the $y$ coordinates reduces (\ref{GintegralP}) to an integral on $M^4$, where the $y$
 dependence of all fields in $L$ disappears because of the delta functions in $\eta$, and
 the ``legs" of $L$ along $dy$ differentials are killed by the product of all independent $dy^{\mu\nu}$ in $\eta$. Thus
 \eq
 S=\int_{M^4} L|_{y=0,dy=0} \label{M4integral}
 \en
is the {\it spacetime action} obtained from the group manifold action (\ref{GintegralP}) with a specific choice of $\eta_{M^4}$. It contains only the usual fields $V^a_\mu(x)$ and $\omega^{ab}_\mu (x)$ of Poincar\'e gravity, and reproduces the first order Einstein-Hilbert action.
 Indeed 
 \begin{align}
 &   \epsilon_{abcd} R^{ab} \we V^c \we V^d |_{y=dy=0} = R^{ab}_{~~\mu\nu} (x) 
 dx^\mu \we dx^\nu \we V^c_\rho (x) dx^\rho  \we V^d_\sigma (x)  dx^\sigma ~\epsilon_{abcd} = \nonumber \\
 &   R^{ab}_{~~ef} (x)  V^e_\mu (x) V^f_\nu (x) ~ dx^\mu  \we dx^\nu  \we V^c_\rho (x) dx^\rho  \we V^d_\sigma (x)  dx^\sigma ~ \epsilon_{abcd} = \nonumber \\
 & R^{ab}_{~~ef} (x)  V^e_\mu (x) V^f_\nu (x)  V^c_\rho (x)   V^d_\sigma (x)  ~ \epsilon_{abcd} ~\epsilon^{\mu\nu\rho\sigma} ~d^4 x=  R^{ab}_{~~ef} (x) ~det V ~ \epsilon^{efcd} \epsilon_{abcd}  ~d^4x
 \nonumber \\
 & = -4~R^{ab}_{~~ab} (x) ~det V~d^4x
 \end{align}
 where the volume 4-form $d^4x$ is defined by $dx^\mu \we dx^\nu \we dx^\rho \we dx^\sigma =
 \epsilon^{\mu\nu\rho\sigma} d^4 x$.
 \sk
 \noi {\bf Note:} $\eta$ is closed (because it contains ``functions" depending on $y$ multiplied by all the $dy$ differentials) and not exact (because of the Dirac deltas $\delta(y)$), and thus belongs  to a nontrivial de Rahm cohomology class. Deformations of
the $M^4$ surface generated by diffeomorphisms leave the Poincar\'e dual $\eta$ in the
same cohomology class, since the Lie derivative commutes with the exterior derivative.
  \sk
 \noi {\bf Discussion}
 \sk
Why go this roundabout way to obtain a well-known gravity action ?
 The answer is at least fivefold:
 
 \noi - all fields have a group-geometric origin, even if they are not all
 gauge fields.
 
 \noi - all symmetries have a common origin as diffeomorphisms on $\Gtilde$, see Section 2.4.
 
\noi  - there is a systematic procedure based on group geometry to construct
  actions, invariant under diffeomorphisms, and under gauge symmetries closing
  on a subgroup of $G$, see Section 5.
  
\noi   - supersymmetry is formulated in a very natural way as a
  diffeomorphism in Grassmann directions of a supermanifold.
  
  \noi - closer contact is maintained with the usual component action, whereas in the superfield formalism the action looks quite different.
  In fact the group manifold action interpolates between the component and the superfield actions of the same supergravity theory, see Section 8.
  
  \subsection{Symmetries}
  
  The action (\ref{GintegralP}) with $\eta$ given in (\ref{eta6form})  is the integral on $\Gtilde$ of a top form: it is clearly invariant
  under diffeomorphisms on $\Gtilde$. But what we are really interested in are the symmetries 
  of the spacetime action as given in (\ref{M4integral}), where the variations are carried out only in the $x$-dependent fields in $L|_{y=0,dy=0}$. The only symmetries guaranteed a priori 
   are the 4-dimensional spacetime diffeomorphisms, the spacetime action being an integral of a 4-form on $M^4$.
  
  Here resides most of the power of the group manifold formalism: if one considers
  the ``mother"  action (\ref{Gintegral}) on $\Gtilde$, the guaranteed symmetries are {\it all} the diff.s on $\Gtilde$, generated by the Lie derivative $\ell_\epsi$ along the tangent vectors  $\epsi = \epsi^A t_A$ of $\Gtilde$.
  But how do these symmetries transfer to the spacetime action ?  
  
  The variation of the group manifold action under diff.s generated by $\ell_\epsilon$ is\footnote{Recall $\ell_\epsi =  \iota_\epsi d + d \iota_\epsi$ so that $\ell_\epsi$(top form) = $d(\iota_\epsi$ top form)}
  \eq
   \delta S = \int_{\Gtilde}  \ell_\epsi (L \wedge \eta )= \int_{\Gtilde}  (\ell_\epsi L) \wedge \eta + L \wedge \ell_\epsi \eta =0
   \en
  modulo boundary terms. One has to vary the fields\footnote{Since $\ell_\epsi$ satisfies the Leibnitz rule,   $\ell_\epsi L$ can be computed by varying in turn all fields inside $L$.} in $L$ as well as the submanifold embedded in $\Gtilde$:  the sum of these two variations gives zero\footnote{In the following the vanishing of action variations will always be understood modulo boundary terms.} on the group manifold action $S$.  
    But what we need
   in order to have a {\it spacetime} interpretation of all the symmetries of $S$, is really
   \eq
   \delta S = \int_{\Gtilde} (\ell_\epsi L) \wedge \eta =0 \label{spacetimesymm}
   \en
 If this holds, varying the fields $\phi$ inside $L$ with the Lie derivative $\ell_\epsilon$ as in (\ref{Lieder}), and then projecting on spacetime
 ($y=0,dy=0$),  yields spacetime variations
 \eq
\de \phi  (y=dy=0) =  \ell_\epsi \phi (x,y) |_{y=dy=0}
 \en
  that leave the spacetime action (\ref{M4integral}) invariant. We call them
 {\it spacetime invariances}. They originate from the diff. invariance of the group manifold action,
 and give rise to symmetries of the spacetime action  (\ref{M4integral}) only when (\ref{spacetimesymm})
 holds. This happens if one of the following conditions is satisfied:
 \sk
 \noi $\bullet$ the Lie derivative on $\eta$ vanishes:
 \eq
 \ell_\epsi \eta = 0 \label{elloneta}
 \en
 \sk
\noi  $\bullet$ the spacetime projection of the Lie derivative of $L$ is exact:
 \eq
 ( \ell_\epsi L )|_{y=dy=0} = d \alpha  \label{ellonL}
  \en
  \noi In this case the variation (\ref{spacetimesymm})
  \eq
  \delta S = \int_{\Gtilde}  (\ell_\epsi L) \wedge \eta =\int_{M^4}    (\ell_\epsi L )|_{y=dy=0}
  \en
  vanishes after integration by parts.
  The requirement  (\ref{ellonL}) is equivalent to 
  \eq
 ( \iota_\epsi dL )|_{y=dy=0} = d \alpha'  \label{idonL}
  \en
 since $\l_\epsi = \iota_\epsi d + d \iota_\epsi$.  
 \sk
 \noi The Lagrangian $L$ depends on the $\Gtilde$-vielbein $\sigma^A$ and its curvature $R^A$, so that also $dL$, after use of Bianchi identities, is expressed in terms of $\sigma^A$ and $R^A$. Then condition (\ref{idonL})  translates
 into a {\it condition on the contractions} $\iota_\epsi R^A$, i.e. a condition on the curvature components.

 Let us see how this works for Poincar\'e gravity. 
  \sk
 \noi {\bf Lorentz gauge transformations}
 \sk
We choose
 $\epsi= \epsi^{ab} t_{ab}$, with $t_{ab}$ tangent vector on $\Gtilde$ dual to $\om^{ab}$, and compute $ \iota_\epsi dL$ in $\Gtilde$ (in $M^4$ we would have trivially $dL=0$ since $L$ is a 4-form). We find\footnote{omitting the symbol $\wedge$ for exterior products between forms.}:
   \eq
 dL = [(dR^{ab}) V^cV^d + 2 R^{ab} (dV^c) V^d  ] \epsilon_{abcd} = 2 R^{ab} R^c V^d  \label{dLPoincare} \epsilon_{abcd}
 \en
 using the Bianchi identities (\ref{bianchiRa}) and (\ref{bianchiRab}). The contraction along
 $\epsi = \epsi^{cd} t_{cd}$ 
 \eq           
 \iota_\epsi dL = 2 (\iota_\epsi R^{ab}) R^c V^d \epsilon_{abcd}+2 R^{ab} (\iota_\epsi R^c) V^d \epsilon_{abcd}
 \en
 \noi  vanishes if the Poincar\'e curvatures satisfy the horizontality conditions
  \eq
   \iota_{t_{cd} }R^a = \iota_{t_{cd}} R^{ab}=0  \label{horizonPoincare}
   \en
 
  In this case $\iota_{\epsi^{cd} t_{cd}} dL=0$ and the spacetime action is invariant under transformations generated by $\ell_{\epsi^{cd} t_{cd}}$.
  The horizontality conditions (\ref{horizonPoincare}) 
 imply that the curvatures have no ``legs" in the Lorentz directions: when expanded on
    a complete basis of 2-forms on $\Gtilde$ as in (\ref{Raexpansion}), (\ref{Rabexpansion}), their $V\omega$ and $\omega\omega$ components vanish  ({\it horizontality} in the Lorentz directions). 
    
    The transformations generated by $\ell_{\epsi^{cd} t_{cd}}$ are found by using the horizontality constraints (\ref{horizonPoincare}) inside the general formula (\ref{Lieder}) and read:
 \eqa
 & & \ell_{\epsi^{cd} t_{cd}} V^a =   \epsi^a_{~b} V^b  \label{LorentzonV}\\
 & & \ell_{\epsi^{cd} t_{cd}}  \omega^{ab} = d \epsi^{ab} - \omega^a_{~c} \epsi^{cb} + \omega^b_{~c} \epsi^{ca}= \Dcal \epsi^{ab} \label{Lorentzonom}
 \ena
They are the usual local Lorentz rotations on the vierbein and the spin connection.

It is easy to check directly the invariance of the action under these transformations, recalling that $R^{ab}= d\om^{ab} - \om^a_{~c} \om^{cb}$ transforms homogeneously under (\ref{Lorentzonom}):
 \eq
 \ell_{\epsi^{cd} t_{cd}} R^{ab}=\epsi^a_{~c} R^{cb} - \epsi^b_{~c} R^{ca} \label{LorentzonRab}
 \en
  
   \sk
   \noi {\bf Note:}   the constraints (\ref{horizonPoincare}) will be derived
   as {\it part} of the equations of motion on $\Gtilde$ in next Section.
   \sk
 \noi   The horizontality constraints can be used {\sl inside} the action (\ref{Gintegral}), i.e. we can consider
   the fields appearing in the action as satisfying the ``partial shell" given by (\ref{horizonPoincare}). 
   Then the soft group manifold $\Gtilde$ takes the structure of a principal fiber bundle with base space 
      $\Gtilde/H$ and fiber $H$, $H$ being the Lorentz group. 
\sk
 \noi {\bf Spacetime diffeomorphisms}
 \sk
 Diff.s along tangent vectors $\partial_\mu$ dual to $dx^\mu$ are known a priori to be invariances of the spacetime action, and we can verify that indeed (\ref{elloneta}) holds, i.e. $\ell_{\epsi^\mu \partial_\mu} \eta_M=0$, since
 $\eta_M$ contains only $dy$ differentials. Diff.s along tangent vectors $t_a$ dual to $V^a$, i.e. generated by $\ell_\epsi$ with
 $\epsi = \epsi^a t_a$ are also spacetime invariances, when one uses the horizontality conditions (\ref{horizonPoincare}). Indeed in this case we find
 $dL=0$ ($dL$ is a 5-form, and cannot contain 5 $V$'s), and therefore also $\iota_{\epsi^a t_a} dL=0$. 
   
 The diff.s along $\epsi=\epsi^a t_a$ act on the fields as:
 \eqa
 & & \ell_\epsi V^a = \Dcal \epsi^a + \iota_\epsi R^a=\Dcal \epsi^a + 2 R^a_{~bc} ~\epsi^b V^c  \\
 & & \ell_\epsi \om^{ab} = \iota_\epsi R^{ab}= 2 R^{ab}_{~~cd} ~ \epsi^c V^d
 \ena
 
 \noi {\bf Note:}  we can verify
   that the horizontality constraints (\ref{horizonPoincare}) are consistent with the Bianchi identities (\ref{bianchiRa}),(\ref{bianchiRab}), by projecting 
the BI on the complete basis of 3-forms $VVV$, $VV\omega$, $V\omega\omega$, $\omega\omega\omega$.

\subsection{Variational principle and field equations}
 
The group manifold action (\ref{Gintegral}) is a functional of $L$ and of the embedded submanifold $M$, and
therefore varying the action means varying both $L$ and $M$. Varying $M$
corresponds to varying $\eta_{M}$. Then the variational principle reads:
\begin{equation}
\label{Svariation}{\ \delta S[L, M] =
\int_{\Gtilde} ( \delta L\wedge\eta_{M} + L\wedge\delta\eta_{M}) } =0\,.
\end{equation}
Any (continuous) variation of $M$ can be obtained by acting on 
$\eta_M$ with a diffeomorphism generated by a Lie derivative $\ell_\xi$.
An arbitrary variation is generated by an arbitrary $\xi$ vector, and
the variational principle becomes
\begin{equation}
\label{Svariation1}{\ \delta S[L, M] =
\int_{\Gtilde} ( \delta L\wedge\eta_{M} + L\wedge\ell_\xi \eta_{M}) } =0\,.
\end{equation}
Since field
variations in $L$ and variation of $M$ are independent, the two terms in (\ref{Svariation1})
must vanish separately. From the vanishing of the first one we deduce
\eq
\int_{\Gtilde} (  \delta \phi \we {\part L \over \part \phi}  +  d \delta \phi \we {\part L \over \part (d\phi) } ) \wedge \eta_M =0
\en
where $L=L(\phi,d\phi)$ is considered a function of the 1-form fields $\phi$ and their ``velocities" $d\phi$.  A summation on all fields is understood.
Integrating by parts and recalling $d \eta_M=0$ yields
 \eq
 \int_{\Gtilde} \delta \phi \we ( {\part L \over \part \phi}  + d {\part L \over \part (d\phi) } ) \wedge \eta_M =0
\en
and since the $\delta \phi$ are arbitrary we find 
 \eq
  ({\part L \over \part \phi} + d {\part L \over \part (d\phi)}) \we  \eta_M=0 \label{ELequations0}
  \en
This must hold for any $\eta_M$ (i.e. for generic embedding functions): we arrive therefore at equations
that hold on the whole $\Gtilde$, and are the form version of
  the Euler-Lagrange equations:
  \eq
{\part L \over \part \phi} + d {\part L \over \part (d\phi)}  =0 \label{ELequations1}
\en
If $L$ is a $d$-form, these equations are $(d-1)$-forms. Their content can be examined by
expanding them along a complete basis of  $(d-1)$-forms in $\Gtilde$.

Requiring the vanishing of the second term in the variation (\ref{Svariation1}) does
not imply further equations besides the Euler-Lagrange field equations (\ref{ELequations1}): indeed this term vanishes on the shell of solutions of Euler-Lagrange equations. To prove it, notice that
\eq
\int_{\Gtilde} L \we \ell_\xi \eta_M = - \int_{\Gtilde} \ell_\xi L \we \eta_M =0 ~(on~shell) \label{onshell}
\en
because $\ell_\xi L$ is just a particular variation of $L$, under which the action remains stationary on-shell.

Thus the group manifold variational principle leads to the field equations (\ref{ELequations1}),
holding as $(d-1)$-form equations on the whole $\Gtilde$. 
\sk
\noi {\bf Note 1:} The variational principle 
{\it does not determine} the embedding of $M$ into $\Gtilde$.
\sk
\noi {\bf Note 2:} the field equations (\ref{ELequations1}) are form equations, and therefore
invariant under the action of a Lie derivative. More precisely, if $\phi$ is a solution of
(\ref{ELequations1}), so is $\phi + \ell_\epsi \phi$: Lie derivatives generate symmetries
of the field equations.

\sk
\noi Finally, we have the following 
\sk
\noi {\bf Theorem:} $dL = 0 ~(on~shell) $
\sk
\noi i.e. the Lagrangian, as a $d$-form on $\Gtilde$, is closed on shell. To prove it
recall that
$\eta_M$ is closed , so that  on shell we find, cf. (\ref{onshell}): 
\eq
0 = \int_{\Gtilde} L \we \ell_\xi \eta_M =  \int_{\Gtilde} L \we d \iota_\xi \eta_M =- (-)^d  \int_{\Gtilde} dL \we \iota_\xi \eta_M
\en
$\xi$ being arbitrary, this implies $dL=0$ (on shell)\footnote{
In fact, this is just Stokes theorem applied to a region of $\Gtilde$ bounded by two different hypersurfaces $M$ and $M'$.} \square

It is interesting to notice
that in many cases $dL= 0$ holds only on a {\it subset} of the equations of motion,
and in some cases, it holds completely off-shell, as we discuss later.
\sk
Let us apply the preceding discussion to the Poincar\'e gravity example.
Varying $\omega^{ab}$ and $V^c$ in the action 
\begin{equation}
S [\omega,V, \eta] = \int_{\Gtilde} R^{ab} V^{c} 
V^{d} \epsilon_{abcd} ~ \eta_{M}
\end{equation}
yields respectively:
\eqa
& & \delta S =\int_{\Gtilde} \Dcal (\delta \om^{ab}) V^c V^d \epsilon_{abcd} ~\eta_M = 
2 \int_{\Gtilde} (\delta \om^{ab}) R^c V^d \epsilon_{abcd} ~\eta_M \\
& & \delta S = 2 \int_{\Gtilde} R^{ab} \delta V^c V^d \epsilon_{abcd}~ \eta_M 
\ena
Imposing $\delta S=0$ leads to the equations of motion 
\eqa
& & R^{c}  V^{d} \epsilon_{abcd}  =0   \label{PoincarefieldeqRa}\\
& & R^{ab}  V^{d} \epsilon_{abcd}  = 0 \label{PoincarefieldeqRab}
\ena
These field equations are 3-form equations on $\Gtilde$. The curvatures
$R^a$ and $R^{ab}$ can be expanded on a complete basis of 2-forms as
\begin{align}
& R^{a} = R^{a}_{~b,c} V^{b} V^c + R^{a}_{~b,cd} V^b  \omega^{cd} +
R^a_{~bc,de} \omega^{bc} \omega^{de} \label{Raexpansion}\\
& R^{ab}=R^{ab}_{~~cd}V^{c}  V^{d} + R^{ab}_{~~c,de}V^{c} 
\omega^{de}+ R^{ab}_{~~cd,ef}\omega^{cd} \omega^{ef} \label{Rabexpansion}
\end{align}
Substituting these expansions into the field equations (\ref{PoincarefieldeqRa}), (\ref{PoincarefieldeqRab}), and projecting 
on a complete basis of 3-forms in $\Gtilde$  yields:
\eqa
& & R^c_{e,f} V^e V^f V^d \epsi_{abcd} + R^c_{e,fg} V^e \omega^{fg}  V^d \epsi_{abcd} + R^c_{ef,gh} \omega^{ef} \omega^{gh} V^d \epsi_{abcd}  =0 \\
& &  R^{ab}_{e,f} V^e V^f V^d \epsi_{abcd} + R^{ab}_{e,fg} V^e \omega^{fg} V^d \epsi_{abcd} + R^{ab}_{ef,gh} \omega^{ef} \omega^{gh} V^d \epsi_{abcd}   =0 
\ena
The three terms in each equation must vanish separately, since $VVV$, $V\om V$,  $\om \om V$ are independent three-forms.

It is easy to see that the $V\om V$ and $\om\om V$ projections of the first equation imply $R^a_{~b,cd} = R^a_{~bc,de} =0$, i.e. horizontality of $R^a$, while the $VVV$ projection yields $R^a_{~b,c} =0$.
Then $R^a$ as a 2-form on $\Gtilde$ must vanish on shell.  From
\eq
R^a_{\mu\nu} =  \part_\mu V^a_\nu - \part_\nu V^a_\mu - \om^a_{~b,\mu} V^b_\nu +
\om^a_{~b,\nu} V^b_\mu = 0
\en
one finds the spin connection in terms of $V$:
\eq
\omega_{ab,\mu}= V^\nu_a V^\rho_b \eta_{cd} ~(\part_{[\mu} V_{\nu]}^c V_\rho^d - \part_{[\mu} V_{\rho]}^c V_\nu^d + \part_{[\nu} V_{\rho]}^c V_\mu^d) \label{omsecondorder}
\en
where $V^\nu_a$ is the inverse vierbein.

Similarly the second field equation implies $R^{ab}_{~c,de} = R^{ab}_{~cd,ef} =0$ (horizontality of $R^{ab}$), and the Einstein
equations for the inner components $R^{ab}_{~~cd}$:
\begin{equation}
R^{ac}_{~~bc} - \frac{1}{2} \delta^{a}_{b} R^{cd}_{~~cd} =0 \label{Einsteineq}
\end{equation}

\noi We can also check that $dL=0$ on a subset of the field equations, since
$dL=2R^{ab} R^c V^d \epsilon_{abcd}$ (cf. (\ref{dLPoincare})) vanishes 
when $R^a=0$, or when using horizontality of the curvatures.
\sk
\noi {\bf Note 3:} The fact that the same horizontality conditions arise both from the request of the spacetime invariance under diffeomorphisms along the Lorentz directions, and from the field equations, should
come as no surprise. Indeed $dL=0$ on shell implies also $\iota_{\epsi^{cd}t_{cd}} dL|_{y=dy=0}=0$ on shell,
so that the field equations imply conditions on the outer components of the curvatures similar to those requested by spacetime invariance. But there is an important difference: the
conditions on the outer components of $R^A$ coming from (\ref{idonL}) must hold off-shell, while those coming from the field equations are on-shell by definition, and may differ
by field equations involving {\sl spacetime components} of the curvatures. 
\sk
\noi {\bf Note 4:} Lorentz gauge invariance of the action is really due to the absence of 
a bare connection $\omega$ in the Lagrangian (\ref{LPoincare}). In this case also the
field equations do not contain bare $\om$ 's, and
their projections with at least one $\omega$ must then contain outer components of the curvatures. Horizontality follows, and Lorentz gauge transformations can be interpreted as diff.s in the Lorentz coordinates $y^{ab}$. 
\sk
\noi {\bf Note 5:} the horizontality constraints arise as outer projections of the equations of motion,
and, as noted in the preceding Section, can be used inside the group manifold action. 
 The corresponding spacetime action (\ref{M4integral}) remains unchanged:  this can be easily verified by substituting $R^{ab}$ with $R^{ab}_{~~cd} V^c V^d$ into (\ref{LPoincare}).

 \sect{Supergravity}
 
 In the group-geometric approach to supergravity theories, the ``big" manifold $\Gtilde$ 
 is a (soft) {\it supergroup} manifold, and there are fermionic vielbeins
$\psi$ (the gravitini) dual to the fermionic tangent vectors in $\Gtilde$.

\subsection{Soft super-Poincar\'e manifold}

The $N=1$ super-Poincar\'e Lie algebra is a superalgebra which extends the algebra given
in (\ref{PoincarePP})-(\ref{PoincareMP}) by means of a spinorial generator $Q_\alpha$
satisfying:
 \eqa
 & & [P_a,\Qbar_\al]=0 \\
 & & [M_{ab},\Qbar_\beta]= -{1 \over 4} \Qbar_\alpha (\gamma_{ab})^\alpha_{~\beta}   \label{sPoincareMQ} \\
 & & \{ \Qbar_\alpha,\Qbar_\beta \} = -i (C\gamma^a)_{\alpha \beta}  P_{a}\label{sPoincareQQ}
  \ena 
$C_{\al\be} $ is the charge conjugation matrix, and the spinorial generator $\Qbar_\alpha  \equiv Q^\beta
C_{\beta\alpha}$ is a Majorana spinor, i.e. $Q^\beta C_{\beta\alpha} = Q^\dagger_\beta (\gamma_0)^\beta_{~\alpha}$.
Thus the super-Poincar\'e manifold has 10 bosonic directions with coordinates $x^a$, $y^{ab}$, parametrizing  translations and Lorentz rotations, and 4 fermionic directions with Grassmann coordinates $\theta^\alpha$, corresponding to the 4 supercharges $\Qbar_\alpha, \alpha=1,..4$. 

The components of the vielbein of the $\Gtilde$ =(soft) superPoincar\'e manifold are the vierbein $V^a$, the spin connection $\omega^{ab}$ and the
gravitino $\psi^\alpha$.
 corresponding respectively to the generators $P_a$, $M_{ab}$ and $\Qbar_\alpha$,

 The curvature (\ref{Gcurvature})
becomes, using the structure constants of the Lie superalgebra:
\eqa
& & R^{a}= dV^{a} - \omega^{a}_{~c} V^{c} - \frac{i }{2} \bar\psi\gamma^{a} \psi \equiv \Dcal V^a  - \frac{i }{2} \bar\psi\gamma^{a} \psi\label{RasuperPoincare}\\
& & R^{ab}=d \omega^{ab} - \omega^{a}_{~c} ~\omega^{cb} \label{RabsuperPoincare}\\
& & \rho= d \psi- \frac{1}{ 4} \omega^{ab} \gamma_{ab} \psi \equiv \Dcal \psi \label{rhosuperPoincare}
\ena
defining respectively the supertorsion, the
Lorentz curvature and the gravitino field strength. $\Dcal$ is the Lorentz covariant exterior derivative.

As a consequence of the definitions (\ref{RasuperPoincare})-(\ref{rhosuperPoincare}), the following Bianchi identities hold:
\eqa
& & dR^a -\omega^a_{~b} R^b + R^a_{~b} V^b - i \psibar \gamma^a \rho \equiv  \Dcal R^a + R^a_{~b} V^b - i \psibar \gamma^a \rho= 0 \label{BianchiRasuperPoincare}\\
& & dR^{ab} - \omega^a_{~c} R^{cb} +  \omega^b_{~c} R^{ca} \equiv \Dcal R^{ab}=0 \label{BianchiRabsuperPoincare}\\
& & d\rho - {1 \over 4} \omega^{ab} \gamma_{ab} \rho + {1 \over 4} R^{ab} \gamma_{ab} \psi \equiv \Dcal \rho 
+ {1 \over 4} R^{ab} \gamma_{ab} \psi =0 \label{BianchirhosuperPoincare}
\ena
\subsection{The supergroup manifold action}

The supergravity action is again the integral of a 4-form on a submanifold $M^4 \in \Gtilde$, diffeomorphic to Minkowski spacetime. In this case $\Gtilde$ is the 14-dimensional superPoincar\'e group manifold, and the action reads: 
\begin{equation}
S[V,\omega,\psi,\eta]= \int_{\Gtilde} (R^{ab} V^{c} V^{d} \epsilon_{abcd} + 4
\bar\psi\gamma_{5} \gamma_{a} \rho V^{a} ) ~\eta_{M^4} \label{SGaction1}
\end{equation}
with  $\eta_{M^4}$ = Poincar\'e dual of $M^4$. Here $\eta_{M^4}$ is a (closed) ``10-form" that localizes
the Lagrangian on the submanifold $M^4$, to be discussed in Section 4 in the context
of integration on supermanifolds. The Lagrangian can be found by use of the building rules 
of Section 5; for a detailed derivation see for ex. \cite{gm21,gm23}.

\subsection{The spacetime action}

The spacetime action is obtained by a specific choice of $\eta_{M^4}$ in (\ref{SGaction1}).
Its precise expression will be given in Section 4.
  Actually a piece of $\eta_{M^4}$ is the 6-form (\ref{eta6form}), that localizes the Lagrangian on $y=dy=0$ 
once the integration on $y$ coordinates is carried out. We will always assume that this integration
has been carried out, so that all fields depend only on $x$ and $\theta$ coordinates. Moreover all curvatures
are taken to be horizontal in the Lorentz directions. As a consequence  the theory lives in a superspace $M^{4|4}$ spanned by four bosonic coordinates $x^a$ and four fermionic coordinates $\theta^\alpha$.

\subsection{Symmetries}

The symmetries of the spacetime action  (spacetime invariances) are those
generated by a Lie derivative $\ell_\epsi$ such that $\iota_\epsi dL|_{\theta=d\theta=0}= d\alpha'$,
cf. (\ref{idonL}). We need to compute $dL$. Using the Bianchi identities 
(\ref{BianchiRabsuperPoincare}) and (\ref{BianchirhosuperPoincare}), and the definition 
of the torsion $R^a$ in (\ref{RasuperPoincare}) we find:
\begin{align}
& dL= 2 R^{ab} R^c V^d \epsi_{abcd} + i R^{ab} \psibar \ga^c \psi V^d \epsi_{abcd}+
4 \rhobar \ga_5 \ga_a \rho V^a + \nonumber \\
&  ~~~~~~~ + \psibar \ga_5 \ga_c \ga_{ab} \psi R^{ab} V^c -4 \psibar \ga_5 \ga_a \rho R^a - 2i \psibar \ga_5 \ga_a \rho \psibar \ga^a \psi \label{dL1}
\end{align}
The gamma identity
\eq
\ga_c \ga_{ab} = \eta_{ac} \ga_b - \eta_{bc} \ga_a +i \epsi_{abcd}\ga_5 \ga^d 
\en
implies $\psibar \ga_5 \ga_c \ga_{ab} \psi =i \epsi_{abcd} \psibar \gamma^d \psi$, so that the second and the fourth term cancel in 
(\ref{dL1}). Moreover from the Fierz identity in Appendix D one deduces
\eq
\gamma_a  \psi \psibar \gamma^a \psi =0 \label{fierz1}
\en
and since $\psibar \gamma_5 \gamma_a \rho = \rhobar \gamma_5 \gamma_a \psi$ also the last term in
(\ref{dL1}) vanishes due to (\ref{fierz1}). Therefore
\eq
dL= 2 R^{ab} R^c V^d \epsi_{abcd} + 4 \rhobar \ga_5 \ga_a \rho V^a 
  - 4 \psibar \ga_5 \ga_a \rho R^a 
\en
\sk
\noi {\bf Lorentz gauge transformations}
\sk
It is immediate to see that if all curvatures are horizontal in the Lorentz directions
(no ``legs" along $\om$) then indeed $\iota_{\epsi^{ab} t_{ab}}dL=0$, and Lorentz
transformations are a spacetime invariance of the supergravity action. This is essentially due to the absence of bare $\omega^{ab}$ in $L$. The general
diffeomorphism formula (\ref{Lieder}) yields the usual Lorentz transformations
 \eqa
 & & \ell_{\epsi^{cd} t_{cd}} V^a =   \epsi^a_{~b} V^b  \label{LorentzonV2}\\
 & & \ell_{\epsi^{cd} t_{cd}}  \omega^{ab} = d \epsi^{ab} - \omega^a_{~c} \epsi^{cb} + \omega^b_{~c} \epsi^{ca}= \Dcal \epsi^{ab} \label{Lorentzonom2} \\
 & &  \ell_{\epsi^{cd} t_{cd}} \psi = {1 \over 4} \epsi^{ab} \gamma_{ab} \psi
 \ena
We can check directly the invariance of the action under these variations: again all curvatures and vierbeins appearing in (\ref{SGaction1}) transform
homogeneously.
\sk
 \noi {\bf Spacetime diffeomorphisms}
 \sk
 Diff.s along tangent vectors $\partial_\mu$ dual to $dx^\mu$ are invariances of the spacetime action, since $\ell_{\epsi^\mu \partial_\mu} \eta_M=0$ due to
 $\eta_M$ containing only $dy$ and $d\theta$ differentials, see Section 4 on superintegration.
 \sk
  \noi {\bf Supersymmetry transformations}
  \sk
  Diff.s along tangent vectors $t_\alpha$ dual to $\psi^\alpha$ are spacetime invariances
  provided $\iota_\epsilon dL|_{\theta=d\theta=0}= total ~derivative$ with $\epsilon = \epsilon^\alpha t_\alpha$, that is to say
  \eqa
 & &   \iota_\epsilon dL = 2 (\iota_\epsilon R^{ab}) R^c V^d \epsi_{abcd} + 2 R^{ab} (\iota_\epsilon R^c ) V^d \epsi_{abcd} + 8 \rhobar \gamma_5 \gamma_a (\iota_\epsilon \rho) V^a \nonumber \\
 & & - 4 \epsilonbar \gamma_5 \gamma_a \rho R^a - 4 \psibar \gamma_5 \gamma_a (\iota_\epsilon \rho) R^a
 - 4 \psibar \gamma_5 \gamma_a \rho (\iota_\epsilon R^a) = tot.~ der. \label{idL2}
   \ena
with $\theta=d\theta=0$.  This is a condition for the contractions on the curvatures, and it is satisfied by:
 \eqa
 & & \iota_\epsilon R^a =0 \label{rh1}\\
 & & \iota_\epsilon R^{ab} = - \epsi^{abef} \rhobar_{ef} \ga_5 \ga_g \epsilon V^g - \epsi^{efg[a} \rhobar_{ef} \ga_5 \ga_g \epsilon V^{b]}  \equiv \thetabar^{ab}_c \epsilon V^c \label{rh2}\\
  & & \iota_\epsilon \rho =0 \label{rh3}
 \ena
Thus we have supersymmetry invariance of the spacetime action if the curvatures have the following parametrization on a basis of 2-forms:
 \eqa
 & & R^a=R^a_{~bc} ~V^b V^c  \label{param1}\\
 & & R^{ab}= R^{ab}_{~~cd} V^c V^d + \thetabar^{ab}_c ~\psi ~V^c  \label{param2}\\
 & & \rho= \rho_{ab}~ V^a V^b \label{param3}
 \ena
 where we have taken into account also horizontality in the Lorentz directions. The conditions (\ref{rh1})-(\ref{rh3}) are called ``rheonomic conditions", and similarly (\ref{param1})-(\ref{param3}) are called
 ``rheonomic parametrizations" of the curvatures.
 
  The diff.s along $\epsilon=\epsilon^\alpha t_\alpha$ (supersymmetry transformations) act on the fields according to the general formula
   (\ref{Lieder}), where the contractions on the curvatures are given in (\ref{rh1})-(\ref{rh3}):
 \eqa
  & & \ell_\epsilon V^a = i \epsilonbar \gamma^a \psi \label{susy1}\\
 & & \ell_\epsilon \om^{ab} =  \thetabar^{ab}_c \epsilon V^c  \label{susy2}\\
 & & \ell_\epsilon \psi = \Dcal \epsilon \equiv d \epsilon - {1 \over 4} \om^{ab} \ga_{ab} \epsilon \label{susy3}
 \ena
with $\thetabar^{ab}_c$ defined in (\ref{rh2}). 

\subsection{Bianchi identities}

The Bianchi identities (\ref{BianchiRasuperPoincare})-(\ref{BianchirhosuperPoincare}) are 
satisfied by the curvatures in (\ref{param1})-(\ref{param3}), where the horizontality and rheonomic conditions
are implemented, provided the following equations on their $VV$ components hold:
\eqa
& & R^a_{~bc}=0 \label{eom1}\\
& & R^{ac} _{~~bc} - {1\over 2} \delta^a_b ~R^{cd}_{~~cd} =0 \label{eom2}\\
& & \gamma^a \rho_{ab}=0\label{eom3}
\ena
i.e. the zero torsion condition that allows to express
$\om$ as function of $V$ and $\psi$, and the Einstein and Rarita-Schwinger
propagation equations for the vielbein and the gravitino, respectively. Thus the Bianchi
identities for the curvatures parametrized as in (\ref{param1})-(\ref{param3}) hold only {\it on the shell of the propagation equations} (\ref{eom1})-(\ref{eom3}). As a consequence the superalgebra generated by
the Lie derivatives closes only on-shell (see Appendix A). In other words, the transformations (\ref{susy1}) - (\ref{susy3}) can be interpreted as diffeomorphisms in $M^{4|4}$ only when they are applied on fields that are solutions of the field equations. On general field configurations the supersymmetry transformations 
(\ref{susy1}) - (\ref{susy3}) leave the action invariant, but their commutator cannot be expressed
as a Lie derivative along a tangent vector of $\Gtilde$. This situation can be cured by adding
extra fields in the theory, called {\it auxiliary fields}, entering in the parametrization of the curvatures
in such a way that the Bianchi identities do not imply propagation equations. The auxiliary fields are nondynamical
fields, but their degrees of freedom (d.o.f.) are needed to ensure an equal number of off-shell d.o.f.
of fermions and bosons. We will see how to achieve this for $d=4$ supergravity in Section 9.

The rheonomic parametrizations (\ref{param1})-(\ref{param3})  cannot be used inside the action, since it would amount to consider only on-shell fields and not all field configurations. They have been used exclusively in the transformation laws of the fields. In fact
they have been {\sl determined} in Section 3.4 by requiring supersymmetry invariance of the spacetime action.

\sk
{\bf Note 1:} horizontality, rheonomic conditions and propagation equations will all be derived in next Section 
as field equations from the group manifold action.
\sk
{\bf Note 2:}  in deriving the rheonomic conditions from (\ref{idL2})
we have tacitly assumed that (\ref{param3}) could be used in the {\sl uncontracted} $\rho$ (and thus outside the expression of a field variation) in the fourth  term of the $\iota_\epsilon dL$ expression. This in fact we can do, since the condition (\ref{idL2}) only needs to hold with $\theta=d\theta=0$, and $\rho=\rho_{\mu\nu} dx^\mu dx^\nu= \rho_{ab} V^a V^b$  when all quantities depend only on $x$ and $dx$.
\sk
{\bf Note 3:} the spacetime action (\ref{SGaction1}) and its invariance under the supersymmetry transformations (\ref{susy1})-(\ref{susy3}) were first found in ref. \cite{FFvN} in second
order formalism and in \cite{DZ} in first order formalism, see also the standard references 
 \cite{PvNreport,FVP} on supergravity.

\subsection{Field equations}

The variational equations for the group manifold action (\ref{SGaction1}) read:
\eqa
& & 2 R^{c}  V^{d} \epsilon_{abcd}  = 0   \label{sPoincarefieldeqRa}\\
& & 2 R^{ab}  V^{c} \epsilon_{abcd} + 4 \psibar \ga_5 \ga_d \rho = 0 \label{sPoincarefieldeqRab} \\
& & 8 \ga_5 \ga_a \rho V^a -4 \ga_5 \ga_a \psi R^a=0 \label{sPoincarefieldeqrho}
\ena
The analysis proceeds as follows. We first expand the curvatures on a basis of 2-forms\footnote{assuming
horizontality in the Lorentz directions. This amounts to consider configurations
satisfying the Lorentz horizontality constraints on the curvatures.}
\eqa
& & R^a = R^a_{~bc} V^b V^c + \thetabar^a_{~c} \psi V^c + \psibar K^a \psi \\
& & R^{ab} = R^{ab}_{~~cd} V^c V^d + \thetabar^{ab}_{~~c} \psi V^c + \psibar K^{ab} \psi \\
& & \rho= \rho_{ab} V^a V^b + H_c \psi V^c + \Omega_{\al\be} \psi^\al \psi^\be
\ena
and then insert them into the field equations (\ref{sPoincarefieldeqRa})-(\ref{sPoincarefieldeqrho}). These, being 3-form equations, can be expanded on the basis $\psi\psi\psi$,  $\psi\psi V$, $\psi VV$, $VVV$. Their content is given below (the three lines correspond to the three eq.s of motion):
\sk
\noi$\psi\psi\psi$ sector:
\eqa
& &\Om_{\al\be} =0\\
& & 0=0\\
& & K^a=0
\ena
$\psi\psi V$ sector:
\eqa
& & 2 \psibar K^{ab} \psi V^c \epsi_{abcd} + 4 \psibar \ga_5 \ga_d H_c \psi V^c=0 \label{psipsiV1}\\
& &  ~~~~~~~~~~~~~~~~~~0=0\\
& & ~~~~~~~~~~~~~~~~~~\thetabar^a_{~c}=0
\ena
$\psi V V$ sector:
\eqa
& & 2 \thetabar^{ab}_{~~e} \psi V^e V^c \epsi_{abcd} + 4 \psibar \ga_5 \ga_d \rho_{ab} V^a V^b =0 \label{psiVV1}\\
& &  ~~~~~~~~~~~~~~~~~~0=0\\
& & \ga_5 \ga_a H_b \psi V^b V^a - 4 \ga_5 \ga_c \psi R^c_{~ab} V^a V^b =0 \label{psiVV3}
\ena
$V V V$ sector:
\eqa
& & R^a_{~bc}=0 \\
& & R^{ac} _{~~bc} - {1\over 2} \delta^a_b ~R^{cd}_{~~cd} =0 \\
& & \gamma^a \rho_{ab}=0
\ena
Inserting $R^a_{~bc}=0$ into (\ref{psiVV3}) yields $H_c=0$, which used in (\ref{psipsiV1})
gives $K^{ab}=0$. Thus the only nontrivial relation in the ``outer" projections is (\ref{psiVV1}),
that determines $\theta^{ab}_{~~c}$ to be
\eq
\theta^{ab}_{~~c}=-\epsi^{abef} \rhobar_{ef} \ga_5\ga_c - \delta^{[a}_c \epsi^{b]efg} \rhobar_{ef} \ga_5\ga_g  \label{thetabar1}
\en
in agreement with the $\theta^{ab}_{~~c}$ obtained from the condition (\ref{rh2}). Thus we arrive at the same curvature parametrizations (\ref{param1})-(\ref{param3})
obtained in Sect. 2 by requiring spacetime supersymmetry invariance.

Finally, the $VVV$ sector reproduces the torsion equation, and the propagation equations
for the vierbein and the gravitino, already obtained from
the Bianchi identities in (\ref{eom1})-(\ref{eom3}).
\sk
\noi{\bf Note:} the torsion equation (\ref{sPoincarefieldeqRa}) has the same solution as in
the pure gravity case:
\eq
R^a=0
\en
with a different definition of $R^a$, cf. (\ref{RasuperPoincare}), that now includes the gravitino field.
In particular 
\eq
2 R^a_{\mu\nu} = \part_\mu V^a_\nu - \part_\nu V^a_\mu - \om^a_{~b,\mu} V^b_\nu +
\om^a_{~b,\nu} V^b_\mu - i \psibar_\mu \ga^a \psi_\nu =0
\en
allows to express the spin connection in terms of $V$ and $\psi$, recovering
second order formalism:
\eq
\om_{ab,\mu} = \omo_{ab,\mu}+{i \over 4} V_a^\nu V_b^\rho (\psibar_\mu \ga_\nu \psi_\rho+
\psibar_\nu \ga_\rho \psi_\mu-\psibar_\rho \ga_\mu \psi_\nu-(\nu \leftrightarrow \rho))
\en
where  $\omo_{ab,\mu}$ is the spin connection of pure gravity in second order formalism, given
in (\ref{omsecondorder}).

\sect{Integration on supermanifolds: integral forms}

We have defined the supergravity action (\ref{SGaction1})  as an integral of a top form on the superPoincar\'e group manifold. We have given explicitly only the 4-form Lagrangian, postponing 
the precise expression of $\eta_M$ to the present Section. In fact in the supergravity case we have tacitly assumed typical properties of bosonic integration, as for ex. the existence of a top form and Stokes' theorem. Here we want to justify these assumptions, and give a short
account of superintegration theory.

The construction of actions invariant under diffeomorphisms is solved 
``ab initio" in ordinary integration theory by {\it form} integration. The integral of a $d$-form
\eq
\omega^{(d)} =\omega_{[\mu_1 \cdots \mu_d]} (x) ~dx^{\mu_1} \wedge \cdots \wedge dx^{\mu_d} \label{dform}
\en
on a $d$-dimensional manifold $M^d$ is defined by
\eq
I =   \int_{M^d} \om^{(d)}  \equiv \int_{M^d} {1 \over d!}~ \omega_{[\mu_1 \cdots \mu_d]} (x) \label{intform}\epsilon^{\mu_1 \cdots \mu_d}  ~d^dx
\en
i.e. by usual (Riemann-Lebesgue) integration on $M^d$ of the function $ {1 \over d!}~ \omega_{[\mu_1 \cdots \mu_d]} (x)\epsilon^{\mu_1 \cdots \mu_d} $, where 
$\epsilon^{\mu_1 \cdots \mu_d}$ is the Levi-Civita antisymmetric symbol in the coordinate basis,
a tensor density of weight $-1$. Therefore
\eq
 \epsilon^{\mu_1 \cdots \mu_d} ~d^dx = \epsilon^{\mu_1 \cdots \mu_d} ~dx^1 \we \cdots \we dx^d
 \en
 is a tensor, and the integrand of (\ref{intform}) is a scalar.
 
 As in the previous Sections, we can consider infinitesimal diffeomorphisms as
 active transformations, generated by the Lie derivative $\ell_\epsilon = \iota_\epsilon d + d \iota_\epsilon $.
 Then the form integral (\ref{intform}) transforms as
 \eq
 \delta I = \int_{M^d} \ell_\epsilon \om^{(d)} =  \int_{M^d} (\iota_\epsilon d + d \iota_\epsilon ) \om^{(d)} =0
 \en
 since $d  \om^{(d)} =0$ ($ \om^{(d)}$ is a top form) and $\int_{M^d} d (\iota_\epsilon \om) =0$ for appropriate boundary conditions. Thus we have checked invariance of the form integral under infinitesimal diff.s
 generated by the Lie derivative. Note that the existence of a top form, namely the fact that a $d$-form is closed on $M^d$, is crucial to ensure action invariance under diff.s.
  
 Can we generalize form integration to {\it supermanifolds}, and use it to construct actions 
 automatically invariant under superdiffeomorphisms ?  The answer to both questions is affirmative.
 
 In analogy 
 with the bosonic case, integration on forms living on supermanifolds is defined via integration of 
functions in superspace. Consider a function $\Phi (x, \theta)$, defined on a supermanifold $M^{d|m}$ 
with $d$ bosonic coordinates $x$ 
and $m$ fermionic (anticommuting) coordinates $\theta^\al$. It is called a {\it superfield}, and can be expanded in the $\theta^\al$ coordinates:
\eq
\Phi(x,\theta)= \phi(x) + \phi_{\alpha_1} (x) \theta^{\alpha_1}+ \phi_{\alpha_1\alpha_2} (x) \theta^{\alpha_1} \theta^{\alpha_2} + \cdots + \phi_{\alpha_1 \cdots \alpha_m} (x) \theta^{\alpha_1} \cdots \theta^{\alpha_m} \label{superfield}
\en
The functions $\phi_{{\al_1} \cdots {\al_p}}(x)$ are called {\it superfield components} ,
and have antisymmetrized indices due to the anticommuting $\theta$'s in the expansion
 (\ref{superfield}). The integral of
the superfield on $M^{d|m}$ is defined by Berezin integration: 
\eq
\int_{M^{d|m}} \Phi (x,\theta) ~d^dx ~d^m \theta \equiv \int_{M^d} {1 \over m!} \phi_{\alpha_1 \cdots \alpha_m} (x)  \epsi^{\alpha_1 \cdots \alpha_m}~ d^dx  \label{Berezin}
\en
Only the highest component of $\Phi$ (corresponding to the maximal number of $\theta$'s) enters the integral on $M^d$.
 
 Note the striking similarity between the two integrals (\ref{intform}) and (\ref{Berezin}). In fact
we can define form integration in terms of Berezin integration. Consider the
 differentials $dx$ in the $d$-form (\ref{dform}) as {\it anticommuting coordinates} $\xi^\mu = dx^\mu$,
 so that $\om^{(d)}$ becomes a {\it function} of $x$ and $\xi$:
 \eq
 \omega^{(d)} (x,\xi) =\omega_{[\mu_1 \cdots \mu_d]} (x) ~\xi^{\mu_1} \cdots  \xi^{\mu_d} 
 \en
Its Berezin integral on $M^{d|d}$ exactly yields the form integral (\ref{intform}). This observation
 is the key for a definition of superform integration on supermanifolds. 
 
A natural generalization of a bosonic top form (\ref{dform}) is a $(d+m)$-superform:
  \eq
  \omega^{(d+m)} (x, \theta)= \omega_{[\mu_1 \cdots \mu_d] \{\alpha_1 \cdots \alpha_m \} } (x, \theta)~
 dx^{\mu_1} \wedge \cdots \wedge dx^{\mu_d} \wedge d\theta^{\alpha_1} \wedge \cdots \wedge d \theta^{\alpha_m}
\label{superform}
 \en
 
\noi  Note that the $d\theta$ differentials {\it commute} (since the $\theta$'s are anticommuting),
  so that the indices $\alpha_i$ are symmetrized. For this reason $\omega^{(d+m)}$ cannot be a top form: a superform can have an arbitrary number of $d \theta$ differentials, and  its exterior derivative does not vanish. Let's ignore for the moment this difficulty, and try to define a superform integral. Inspired by 
the observation in the preceding paragraph, we consider the superform $\omega^{(d+m)}(x,\theta)$
as a function of $x,\theta,dx,d\theta$, i.e. a function of the commuting variables $x, d\theta$ and
anticommuting variables $\theta, dx$. Its integral can be defined by Berezin integration on
 $\theta,dx$, and usual Riemann-Lebesgue integration on $x,d\theta$. Here a second difficulty arises:
 the ordinary integration on the $u= d\theta$ coordinates produces integrals of the type 
 \eq
 \int u^m ~d^m u
 \en
and there is no algorithmic way to assign a $C$-number to it.
For the integral on the even variables $u=d\theta$ to make sense, the integrand must have compact support as a function of $u$.  For this reason we consider functions of the $d\theta$'s which are {\it distributions} in $d\theta$ with support at the origin:
\eq
\omega (x,\theta,dx,d\theta)= \omega_{[\mu_1 \cdots \mu_d]} (x,\theta)~ dx^{\mu_1} \cdots dx^{\mu_d} \delta(d\theta^{1}) \cdots \delta(d\theta^{m})
\en
These ``functions" can be integrated on the supermanifold $M^{d+m|d+m}$
spanned by the $d+m$ bosonic variables $x,d\theta$ and $d+m$ fermionic variables $dx,\theta$.
The integral
\eq
\int_{M^{d+m|d+m}} \omega (x,\theta,dx,d\theta) ~d^dx~ d^m\theta ~d^d(dx) ~d^m (d\theta)
\en
is defined by Berezin integration on the odd variables $dx,\theta$ and usual Riemann-Lebesgue 
integration on the even variables $x,d\theta$. Carrying out integration on the variables $dx$ and $d\theta$ the
integral becomes
\eq
\int_{M^{d|m}} \omega_{[\mu_1 \cdots \mu_d]} (x,\theta) \epsilon^{\mu_1 \cdots \mu_d} ~d^dx ~d^m \theta
\label{intdm}
\en
 This integral can also be seen as
an integral of the {\it form}:
\eq
\omega^{d|m} = \omega_{[\mu_1 \cdots \mu_d]} (x,\theta)
\delta(u^{1}) \cdots \delta(u^{m})~ dx^{\mu_1} \we \cdots \we dx^{\mu_d} \we du^{1} \we\cdots \we du^{m} \label{inttop}
\en
where the even variables $u$ are the differentials $d\theta$.  Indeed, let us integrate this form with 
the recipe of considering it a function of $x,\theta,u$ and of the differentials $dx$,$du$, and then using Berezin and Riemann integration according to the odd or even grading of the variables. The result coincides with  (\ref{intdm}).

Thus the form $\omega^{d|m}$ can be integrated, even if it contains $d\theta$ differentials. 
We achieve this by confining the $d\theta$'s inside delta functions, and in this way overcome the first difficulty 
encountered with the superforms (\ref{superform}). But can $\omega^{d|m}$ overcome also the
second difficulty, and be a {\it top form}? The answer is yes: the $dx$ and $du$ differentials are all anticommuting, so that their number
in $\om^{d|m}$ is already maximal, and multiplying it by $d\theta$ differentials gives zero because
of the presence of the deltas. Therefore $d\om^{d|m}=0$, and $\om^{d|m}$ is a {\it bona fide} top form.
Since it can be integrated and it is a top form, $\om^{d|m}$ is called an {\it integral top form}.

Finally, using the notation
\eq
\delta(u^{1}) \cdots \delta(u^{m})~ du^{1} \we\cdots \we du^{m} \equiv \delta(u^{1}) \we \cdots \we \delta(u^{m})~ 
\en
the integral top form can be rewritten (using $u=d\theta$):
\eq
\omega^{d|m} = \omega_{[\mu_1 \cdots \mu_d]} (x,\theta)
~ dx^{\mu_1} \we \cdots \we dx^{\mu_d} \we \delta(d\theta^{1}) \we \cdots \we \delta(d\theta^{m}) \label{inttop2}
\en
or also
\eq
\omega^{d|m} = \omega_{[\mu_1 \cdots \mu_d] [\al_1 \cdots \al_m]} (x,\theta)
~ dx^{\mu_1} \we \cdots \we dx^{\mu_d} \we \delta(d\theta^{\al_1}) \we \cdots \we \delta(d\theta^{\al_m}) \label{inttop3}
\en
where indices $\al$ are antisymmetrized since the $\delta (d\theta^\al)$ anticommute, and
\eq
m! ~\omega_{[\mu_1 \cdots \mu_d]} \equiv \omega_{[\mu_1 \cdots \mu_d] [1 \cdots m]}
\en
In this notation $\mu$ and $\al$ indices play a similar role, and are both antisymmetrized. The numbers
$d,m$ are respectively called the {\it form number} and the {\it picture number}, and for integral top forms
they coincide with the numbers of bosonic and fermionic dimensions of the supermanifold $M^{d|m}$.

We call ``superforms" the forms of the kind (\ref{superform}), with $dx$ and $d\theta$ differentials, 
without $\delta (d\theta)$'s. Thus superforms have a form number that counts the $dx,d\theta$
differentials, and zero picture number. For example the Lagrangian in (\ref{SGaction1}) is a 
superform $L^{4|0}$. 

\sk
\sk
\noi {\bf Integration on submanifolds of supermanifolds}
\sk
Supergravity actions on supergroup manifolds $\Gtilde$ are given by integrals of a $d$-form Lagrangian $L$ on
a $d$-dimensional (bosonic) submanifold $M^d$ of $\Gtilde$. They can be written as integrals on the whole
$\Gtilde$ of the Lagrangian multiplied by an appropriate Poincar\'e dual $\eta_{M^d}$ of $M^d$, such that
$L \we \eta_{M^d}$ becomes an integral top form. Let us see how this works for $N=1$, $d=4$ supergravity.

The supergravity Lagrangian in (\ref{SGaction1}) is a $(4|0)$ superform. For simplicity we now assume
that fields satisfy the Lorentz horizontality constraints on all the curvatures, and thus effectively
depend only on the superspace coordinates $x^\mu, \theta^\alpha$, with $\mu=1,..4$, $\al=1,..4$.
Then $\Gtilde$ is $M^{4|4}$ superspace, and only integral top forms of type $(4|4)$ can be 
integrated on $M^{4|4}$. We therefore need a Poincar\'e dual of type $(0|4)$, so that
\eq
L^{4|0} \we \eta_{M^4}^{0|4} 
\en
is an integral top form, i.e. of type $(4|4)$. For this purpose we choose:
\eq
\eta_{M^4}^{0|4} = \epsi_{\al\be\ga\de} \theta^\al \theta^\be \theta^\ga \theta^\de ~ \epsi_{\al'\be'\ga'\de'}
\delta (d\theta^{\al'}) \we \delta (d\theta^{\be'}) \we \delta (d\theta^{\ga'}) \we \delta (d\theta^{\de'})
\en
so that
\eq
\int_{M^{4|4}} L^{4|0} \we  \eta_{M^4}^{0|4} = \int_{M^{4}} L^{4|0} (\theta=0, d\theta=0)
\en
and we obtain a spacetime action, where all fields depend only on $x$-coordinates (the terms containing 
$\theta$'s are annihilated by the presence of the 4 $\theta$'s in $\eta$) and have no ``legs"  
$d\theta$ because of the $\delta (d\theta)$ in $\eta$. Note that $\eta_{M^4}^{0|4}$ is closed,
and the explicit $\theta$'s prevent it to be exact.

Since multiplying by the Poincar\'e dual changes the picture number of the resulting form,
$\eta$ is also called Picture Changing Operator (PCO), a name borrowed from string theory
and string field theory.

The Poincar\'e dual is by no means unique: we can orient the $M^4$ surface inside $\Gtilde$
in many different ways. For example consider the PCO obtained by acting on $\eta$ with an infinitesimal diffeomorphism in the $\theta$ directions:
\eq
\eta ' = \eta + \ell_\epsilon \eta = \eta + d(\iota_\epsilon \eta)  \label{etachange}
\en
This is still a PCO, being closed and not exact\footnote{because $\eta$ is closed and not exact, and 
$d$ commutes with $\ell_\epsilon$.}, and dual to a submanifold diffeomorphic to the original $M^4$.
Note also that the change in $\eta$ is exact.

 \sect{Building rules}
 
 \subsection{The Lagrangian $d$-form}
 
 The group geometric approach provides a systematic set of building rules \cite{gm21} for
 constructing Lagrangians of supersymmetric theories:
 \sk
 1) Choose a Lie (super)algebra $G$, containing generators $P_a$ that can be
 associated to $d$ spacetime directions, and a Lorentz-like subalgebra $H$.
Examples are the superPoincar\'e algebras in $d$ dimensions or their uncontracted versions
 (orthosymplectic superalgebras $OSp(N|2^{[d/2]})$). The fields of the theory
 are the vielbein components of the soft group manifold $\Gtilde$.
 \sk
 2) Construct the most general $d$-form on $\Gtilde$, by multiplying (with exterior products) 1-form  vielbein components $\sigma^A$ and 2-form curvatures $R^A$, without bare Lorentz connection and contracting indices with $H$-invariant tensors, so that the resulting Lagrangian 
 is a Lorentz scalar.  
 \sk
 3) Require that the variational equations admit the ``vacuum solution"  $R^A=0$, described by the
 vielbein of the rigid group manifold $G$.
 \sk
4) The construction is greatly helped by scaling properties of the fields, dictated by the structure
 of the Lie (super)algebra $G$, or equivalently by the Cartan-Maurer equations for the $G$ vielbein. Consider for example the superPoincar\'e algebra: it is invariant under the rescalings $P_a \rightarrow \lambda P_a, M_{ab} \rightarrow M_{ab}, \Qbar_\alpha \rightarrow \lambda^{1\over 2} \Qbar_\alpha$. Then the curvature definitions (\ref{RasuperPoincare})-(\ref{rhosuperPoincare}) are invariant under 
 \eq
 V^a \rightarrow \lambda V^a, ~~~\om^{ab} \rightarrow \om^{ab}, ~~~\psi \rightarrow \lambda^{1\over 2} \psi \label{rescalings1}
 \en
 The field equations must be invariant under these rescalings, and therefore the action must scale homogeneously under (\ref{rescalings1}). Since the Einstein-Hilbert term scales as $\lambda^2$, all terms must scale in the same way, and this restricts the candidate terms in the Lagrangian. 
 \sk
 5) Finally, requiring
 that all terms have the same parity as the Einstein-Hilbert term further narrows the list of candidates.
 \sk
 Following the above rules, one arrives at the  $d=4$ supergravity action (\ref{SGaction1}), see for ex.
 \cite{gm21} for a detailed derivation.
 
 \subsection{The use of Bianchi identities}
 
  We have seen in the preceding Sections how a geometrical theory can be 
  constructed, and its action found, starting from a Lie (super)algebra $G$. 
  
  In many cases, however,  the field equations of the theory and their invariances can be  
 derived directly {\it without reference to an action}, using only the Bianchi identities
 and rheonomic constraints on the curvatures.  As discussed 
 in Section 3.5, the Bianchi identities\footnote{The Bianchi identities are identities only if the curvatures are not constrained. }  imply the field equations of $N=1$, $d=4$ supergravity
 when the rheonomy and horizontality constraints hold on the outer components of the curvatures.
  
The rheonomy constraints were deduced by requiring supersymmetry invariance of the spacetime action. 
How can we find them without an action ?  

We first observe that some constraints are needed on the curvatures: their outer components must not contain new fields, besides the ones present in the spacetime theory, to avoid unwanted new degrees of freedom. Then the outer components must be expressed in terms of the inner components, which are the ones involved in the spacetime theory. Implementing these constraints into the Bianchi identities determines the exact form of the outer components and, as we have seen in Section 3.5, may imply conditions also on the inner components. These extra conditions are the propagation equations.
\sk
\noi In the $N=1$, $d=4$ supergravity case, the procedure runs as follows: 
\sk
 1) first expand the
 (soft) group manifold curvatures on a basis of 2-forms as
\eq
R^A=R^A_{ab} V^a V^b +R^A_{\alpha b} \psi^\alpha V^b + R^A_{\alpha \beta} \psi^\al \psi^\beta
\en

 2) then require that the outer components $R^A_{\alpha b}, R^A_{\alpha \beta}$ be expressed only in terms of the inner ones $R^A_{ab}$. This is the requirement of ``rheonomy", and together
 with the scaling properties of the fields and curvatures determines most of the structure of the outer components. 
 \sk
3) Finally, inserting the rheonomic parametrizations of the curvatures into the Bianchi identities yields the precise form (and fixes coefficients) of the outer components, and moreover produces the
equations for the inner components, i.e. the propagation equations.
\sk
This procedure yields a result for the outer components of the curvatures {\it different} from the one obtained in (\ref{param1})-(\ref{param3}) by requiring 
action invariance. The difference resides
 in the expression for
$\thetabar^{ab}_c$.  From the Bianchi identities we find
\eq
\thetabar^{ab}_c = 2i \rhobar_c^{~[a} \ga^{b]} - i \rhobar^{ab} \ga_c \label{thetabarBianchi}
\en
which differs from the $\thetabar^{ab}_c$ found in (\ref{rh2}) or in (\ref{thetabar1}) by a term proportional to the gravitino propagation equation.

As a consequence the supersymmetry variations obtained from eq. (\ref{Lieder}) by using the $\thetabar^{ab}_c$ in (\ref{thetabarBianchi}) differ
in the spin connection variation (\ref{susy2}). Since the difference is proportional to a field equation, they are still invariances of the equations of motion.

\subsection{On shell and off shell degrees of freedom}

In most examples of supersymmetric theories there is a matching between
bosonic and fermionic degrees of freedom. Therefore the choice of the starting super Lie algebra (or free differential algebra, see next Section) 
must take this matching into account. We summarize the counting of d.o.f. in Table 1, where $V_\mu^a$ is the vielbein, $\psi_\mu^\al$ a real (Majorana) gravitino, $\lambda$ a real spinor, $A_\mu$ a gauge vector,  $A_{[\mu_1 \cdots \mu_p]}$ an antisymmetric tensor (components of a $p$-form). If the spinors are complex their d.o.f. are doubled, and if they are Majorana-Weyl their
d.o.f. are halved. The spacetime dimensions and signatures for Majorana-Weyl fermions are discussed in Appendix B.
\sk
The counting of on-shell d.o.f. summarized in Table 1 is obtained by recalling that:
\sk
\noi- only transverse components contribute to on-shell d.o.f. ($d \rightarrow d-2$).

\noi - the dimension of the spinor representation in $d$ dimensions is $2^{[d/2]}$, where
$[~]$ indicates the integer part, and the Dirac equation reduces
the d.o.f by a factor $1/2$.

\noi - for the vielbein $V^a_\mu$ Lorentz gauge invariance reduces the d.o.f. to those of a symmetric tensor.
Taking into account transversality and subtracting the spinless trace gives 
\eq
{(d-2)(d-1)\over 2} -1= {d(d-3) \over 2}
\en
\noi - in the case of the spin $3/2$ gravitino $\psi_\mu^\al$ the gauge condition $\ga^\mu \psi^\al_\mu =0$
eliminates the spin $1/2$ part. Hence the coordinate index $\mu$ contributes with a factor $(d-2)-1$ to the
spinorial d.o.f.
\sk
The counting of off-shell d.o.f. is obtained by subtracting only gauge invariances, without
using equations of motion. Then $d \rightarrow d-1$ for coordinate indices, and no halving
occurs in spinorial d.o.f. counting.

\begin{table}[htbp]
\caption{Off-shell and on-shell degrees of freedom}
\centering
\sk
\begin{tabular}{c c c }
\hline\hline
field & off-shell d.o.f. & on-shell d.o.f.\\ [0.5ex] 
\hline
& & \\
$V_\mu^a$ & ${d(d-1)\over 2}$ &${d(d-3) \over 2}$\\
& & \\
$\psi_\mu^\alpha$& $(d-1)~2^{[d/2]}$ &$ {1 \over 2} (d-3) ~ 2^{[d/2]}$ \\
& & \\
$\lambda^\alpha$& $2^{[d/2]}$ &$ {1 \over 2} ~ 2^{[d/2]}$ \\
& & \\
$A_\mu$&$d-1$&$d-2$  \\
& & \\
$A_{[\mu_1 \cdots \mu_p]} $ & $
\left(
\begin{array}{c}
 d-1 \\
  p 
\end{array}
\right)
$
&  $
\left(
\begin{array}{c}
 d-2 \\
  p 
\end{array}
\right) $
 \\ 
 & & \\
\hline
\end{tabular}
\label{table:dof}
\end{table}

  \sect{Free differential algebras}
  
The dual formulation of Lie algebras provided by the Cartan-Maurer 
equations (\ref{CM}) can be naturally extended to $p$-forms ($p >1$):
\eq
d \si^i_{(p)}+\sum {1 \over n} C^i_{~i_1...i_n} 
\si^{i_1}_{(p_1)}\we ...\we \si^{i_n}_{(p_n)}=0,~~~~~
p+1=p_1+...+p_n  \label{FDA}
\en
\noi where $p, p_1,...p_n$ are, respectively, the degrees of the forms 
$\si^i, \si^{i_1},..., \sigma^{i_n}$, the indices 
$i, i_1,..., i_n$ run on 
irreps of the (super)group $G$, and $C^i_{~i_1...i_n}$ are generalized structure 
constants satisfying generalized Jacobi identities\footnote{obtained by applying 
$d$ to (\ref{FDA}) and requiring $d^2=0$.}. When 
$p=p_1=p_2=1$ and $i, i_1, i_2$ belong to the adjoint representation of 
$G$, eq.s (6.1) reduce to the ordinary Cartan-Maurer equations. The 
(anti)symmetry properties of the indices $i_1,...i_n$ depend on the 
bosonic or fermionic character of the forms $\si^{i_1},...\si^
{i_n}$

If the generalized Jacobi identities hold, eq.s (\ref{FDA})  define a {\sl free 
differential algebra} \cite{sullivan,DFd11,gm3,gm21} (FDA). 
The possible FDA extensions $G'$ of a Lie 
algebra $G$ have been studied in ref.s \cite{sullivan,gm3,gm21}, and rely 
on the existence of 
Chevalley cohomology classes in $G$ \cite{CE}. Suppose that, given an ordinary 
Lie algebra $G$, there exists a $p$-form:
\eq
\Omega^i_{~(p)} (\sigma)=\Omega^i_{~A_1...A_p} \sigma^{A_1} \we ...\we
\sigma^{A_p},~~{\rm \Omega^i_{~A_1...A_p} =constants,~i~runs~on~a~G-
irrep} 
\en
\noi which is covariantly closed but not covariantly exact, i.e.
\eq
\nabla \Omega^i_{~(p)} \equiv d\Omega^i_{~(p)} + \sigma^A \we D(T_A)^
i_{~j} \Omega^j_{~(p)}=0,~~~~\Omega^i_{~(p)} \not= \nabla \Phi^i_{~(p-1)
}  \label{cohomology}
\en
\noi Then $\Omega^i_{~(p)}$ is said to be a representative of a 
Chevalley cohomology class in the $D^i_{~j}$ irrep of $G$. $\nabla$ 
is the boundary operator satisfying $\nabla^2=0$ ($\nabla^2$ would be 
proportional to the curvature 2-form on the {\sl soft} group manifold).
The existence of $\Omega^i_{~(p)}$ allows the extension of the original 
Lie algebra $G$ to the FDA $G'$:
\eqa
& & d \sigma^A+\unmezzo \CABC \sigma^B \we \sigma^C=0  \nonumber\\
& &  \nabla \sigma^i_{(p-1)} + \Omega^i_{(p)} (\sigma)=0  \label{FDA1}
\ena
\noi where $\sigma^i_{(p-1)}$ is a new $p-1$-form, not contained in $G$. 
Closure of eq.s (6.4) is ensured because $\nabla \Omega^i_{(p)} = 0$.

It is clear that $\Omega^i_{(p)}$ differing by exact pieces $\nabla \Phi
^i_{(p-1)}$ lead to equivalent FDA's, via the redefinition $\sigma^i_{(p
-1)} \rightarrow \sigma^i_{(p-1)} + \Phi^i_{(p-1)}$. What we are 
interested in are really nontrivial cohomology classes satisfying eq.s
(6.3). 

The whole procedure can be repeated on the free differential algebra $G'$ 
which now contains $\sigma^A$, $\sigma^i_{(p-1)}$. One looks for the 
existence of polynomials in $\sigma^A$, $\sigma^i_{(p-1)}$

$$\Omega^i_{(q)} (\sigma^A, \sigma^i_{(p-1)})= \Omega^i_{A_1...A_r i_1...i_s} \sigma
^{A_1} \we ... \we \sigma^{A_r} \we \sigma^{i_1}_{(p-1)} \we ... \we 
\sigma^{i_s}_{(p-1)}  \label{polyFDA}$$

\noi satisfying the cohomology conditions (\ref{cohomology}). If such a polynomial 
exists, the FDA of eq.s ({\ref{FDA1}) can be further extended to $G''$, and so 
on.

In constructing $d$-dimensional supergravity theories we usually choose 
as starting superalgebra $G$ the superPoincar\'e Lie algebra, whose Cartan-Maurer
equations can be read off the curvature definitions in 
eq.s (\ref{RasuperPoincare})-(\ref{rhosuperPoincare}). 
The possible $G'$ extensions to 
FDA's depend on the spacetime dimension $d$. For 
example in $d=11$ there 
is a cohomology class of the superPoincar\'e algebra in the identity 
representation:
\eq
\Omega (V,\om,\psi)=\unmezzo \psib  \Ga^{ab} \psi  V^a  V^b 
\en
\noi $d\Omega=0$ holds because of the $d=11$ Fierz identity
\eq
\psib  \Ga^{ab} \psi ~ \psib  \Ga_a \psi =0 
\en
\noi This allows the extension of the algebra (3.4) by means of a three-form $A$:
\eq
dA-\Omega (V,\om,\psi)=0  \label{Acurvature}
\en
\noi {\bf Note 1:} only nonsemisimple algebras can have FDA extensions in 
nontrivial $G$-irreps. Indeed a theorem by Chevalley and Eilenberg \cite{CE}
states that there is no nontrivial cohomology class of $G$ in 
nontrivial $G$-irreps when $G$ is semisimple.
\sk
As for ordinary Lie algebras, we find a 
dynamical theory based on FDA's by allowing nonvanishing curvatures.
This means, for example, that $d=11$ supergravity is based on a 
deformation of the fields $V,\om,\psi,A$ such that the superPoincar\'e 
curvatures and the $A$-curvature defined by the l.h.s. of of (\ref{Acurvature})  are different from zero.
The construction of the action proceeds along the same lines of 
Section 3, and we refer the reader to ref. \cite{gm21} for an exhaustive 
treatment.
\sk
\noi The next two Sections provide examples of FDA's in $d=3$ and $d=4$.  Other theories 
containing higher forms and obtained as gaugings of free differential algebras can be
found in  \cite{d6,d10-0,d10-1,gm21,d10-2}.

\sk
\noi {\bf Note 2:} a ``resolution" of $FDA$'s in terms of larger Lie (super)algebras,
by expressing the $p$-forms with $p > 1$ as products of 1-form fields involving 
new fields, has been considered already in the seminal reference \cite{DFd11} for $d=11$ supergravity. Recent developments of this idea can be found in ref.s \cite{FDAnew1,FDAnew2,FDAnew3}.
\sk

\noi {\bf Note 3:} a dual formulation of FDA's, based on a generalized Lie derivative ``along antisymmetric tensors" has been developed in ref.s \cite{FDAdual1,FDAdual2,FDAdual3,FDAdual4} and leads to nonassociative extensions of Lie (super)algebras.

\sect{Off-shell $N=1, d=3$ supergravity}

Three dimensional supergravity is one of the simplest models of a consistent extension of
general relativity that includes fermions and local supersymmetry. 
 The superfield action (see for ex. \cite{GGRS,RuizPvN}), supplemented by {\it ad hoc} constraints consistent with the Bianchi identities, provides an off-shell formulation of $d=3$ supergravity, local supersymmetry being
realized as a diffeomorphism in the fermionic directions.

On the other hand, the construction of off-shell $d=3$, $N=1$ supergravity in the group geometric
approach \cite{if3} provides an action which
yields both the correct spacetime equations of motion, {\it and} the constraints on the curvatures. The action is written as a
Lagrangian 3-form integrated over a bosonic submanifold of a supermanifold $M^{3|2}$.

As discussed in \cite{if3}, the same action
can be written as an integral over the whole supermanifold of an integral form,
using the Poincar\'e dual that encodes the embedding of the 3-dimensional bosonic submanifold, see Section 8.

\subsection{Off shell degrees of freedom}

The theory contains a vielbein 1-form $V^a$ with 3 off-shell degrees of freedom ($d(d-1)/2$ in $d$ dimensions), and a gravitino $\psi^\al$ with 4 off-shell degrees of freedom ($(d-1)2^{[d/2]}$ in $d$ dimensions for Majorana or Weyl). The mismatch can be cured by an extra bosonic d.o.f., here provided by a bosonic 2-form auxiliary field $B$.

\subsection{The extended superPoincar\'e algebra}

The algebraic starting point is the FDA that enlarges the $d=3$ superPoincar\'e Cartan-Maurer equations
to include the auxiliary 2-form field $B$. This extension of the superPoincar\'e algebra is
possible due to the existence of the $d=3$ cohomology class $\Omega= \psibar \gamma_a \psi V^a$,
closed because of the $d=3$  Fierz identity (\ref{Fierz3d}).

The FDA yields the definitions of the Lorentz curvature, the torsion, the gravitino field strength and the 2-form field strength:
   \eqa\label{d3cm1}
  & & R^{ab}=d \omega^{ab} - \omega^a_{~c} ~ \omega^{cb} \\
   & & R^a=dV^a - \omega^a_{~b} ~ V^b - {i \over 2} \psibar \gamma^a \psi \equiv \Dcal V^a - {i \over 2} \psibar \gamma^a \psi\ \label{torsionRa}\\
   & & \rho = d\psi - {1 \over 4} \omega^{ab} \gamma_{ab} ~ \psi \equiv  \Dcal \psi  \label{d3cm3}\\
   & & H=dB-{i \over 2} \psibar \gamma^a \psi ~V^a    \ena
where $\Dcal$ is the Lorentz covariant derivative. The generalized Cartan-Maurer equations are invariant under the rescalings
     \eq
     \omega^{ab} \rightarrow \lambda^0 \omega^{ab}, ~V^a \rightarrow \lambda  V^a,~\psi \rightarrow \lambda^{1\over 2} \psi,~B \rightarrow \lambda^2 B \label{rescalings3}
           \en
     Taking exterior derivatives of both sides yields the Bianchi identities:
     \eqa\label{parC}
    & &  \Dcal R^{ab} =0 \\
    & &  \Dcal R^a + R^a_{~b} ~ V^b - i~ \psibar \gamma^a \rho =0\\
    & & \Dcal \rho + {1 \over 4} R^{ab} \gamma_{ab} ~\psi =0\\
    & & dH- i~ \psibar \gamma^a \rho V^a + {i \over 2} \psibar \gamma^a \psi~ R^a = 0
     \ena
     
          \subsection{Curvature parametrizations}

As explained in Section 3, the redundancy introduced by promoting each physical field to a superfield has to be tamed by imposing some algebraic constraints on the curvature parametrizations. They are known as {\it 
conventional constraints} in the superspace language and as {\it rheonomic parametrizations} in the 
group manifold approach. Carrying out the protocol of Section 5.2, we find the following parametrizations
       \eqa
         & & R^{ab} = R^{ab}_{~~cd} ~V^c V^d + \thetabar^{ab}_{~~c}~\psi ~V^c + c_1~ f ~\psibar \gamma^{ab} 
         \psi \label{parRab} \\
        & & R^a = 0  \label{parRa}\\
        & & \rho = \rho_{ab} V^a V^b + c_2~f~\gamma_a \psi ~V^a \\
        & & H = f~V^a V^b V^c \epsilon_{abc} \\
        & & df = \partial_a f~ V^a + \psibar \Xi \label{pardf}
        \ena
        with
         \eqa
         \thetabar^{ab}_{~~c, \al} = c_3 ~(\bar\rho_c^{~[a} \gamma^{b]})_\al + c_4 (\rhobar^{ab} \gamma_c)_
         \al~,~~~~~~~~~
         \Xi^\al =c_5 ~ \epsilon^{abc} (\gamma_a \rho_{bc})^\al
         \ena
         The coefficients $c_1,c_2, c_3, c_4,c_5$ are fixed by the Bianchi identities
        to the values:
         \eq
         c_1=  {3i \over 2} ,~c_2= {3 \over 2} ,~c_3 = 2i,~c_4=-i,~c_5 = - {i \over 3!}
         \en
         The $VVV$ component $f$ of $H$ scales as $f \rightarrow \lambda^{-1} f$,
         and is identified with the auxiliary scalar superfield of the superspace approach of ref \cite{RuizPvN}.
         Note that, thanks to the presence of the auxiliary field, the Bianchi identities do not imply
         equations of motion for the spacetime components of the curvatures.
         
         \subsection{The Lagrangian}

Applying the building rules of Section 5 yields the Lagrangian 3-form
          \eq
      L^{3|0} =  R^{ab} V^c \epsilon_{abc} + 2i \psibar \rho + \alpha (f H - {1 \over 2} f^2 V^a V^b V^c \epsilon_{abc})
           \label{d3lagrangian}
            \en
It is obtained by considering the most general Lorentz scalar 3-form, given in terms of the FDA curvatures
            and fields, invariant under the rescalings discussed above, and such that
        the variational equations admit the vanishing curvatures solution
             \eq
             R^{ab} = R^a = \rho=H = f = 0\,,
             \en
           The remaining parameter is fixed to $\alpha = 6$ by requiring $\iota_\epsilon dL^{3|0} = 0$ up to exact terms, i.e. supersymmetry invariance of
           the spacetime action, cf. (\ref{idonL}). In fact with $\alpha=6$ 
           we find
           \eq
           dL^{3|0}=0
           \en
           on the (off-shell) field configurations satisfying the curvature parametrizations (\ref{parRab})-(\ref{pardf}).
           
     \subsection{Off-shell supersymmetry transformations}
       
           The off-shell closure of the supersymmetry transformations
           is ensured because the Bianchi identities hold without recourse to the 
           spacetime field equations. The action is invariant under these transformations, given by the Lie derivative of the fields along  the fermionic directions:
                \eqa
              & &  \delta_\epsi V^a = -i \psibar \gamma^a \epsi \\
              & &  \delta_\epsi  \psi = \Dcal \epsi + {3 \over 2} f~ \gamma_a \epsilon V^a\\
              & & \delta \omega^{ab}= \thetabar^{ab}_{~~c} ~ \epsi V^c - 3i f~ \psibar \gamma^{ab} \epsi \\
              & & \delta_\epsi  B = - i \psibar \gamma^a \epsi V^a \\
              & & \delta_\epsi f = \epsilonbar ~ \Xi
                \ena
               and closing on all the fields without need of imposing the field equations.
               
     \subsection{Field equations}
  
   Varying $\omega^{ab}$, $V^a$, $\psi$, $B$ and $f$ leads to the equations of motion:
                \eqa \label{reoEQ}
                 & & R^a=0 \\
                 & & R^{ab} = 9 f^2 ~V^a V^b + {3i \over 2} f ~ \psibar \gamma^{ab} \psi \\
                 & & \rho = {3 \over 2} f ~\gamma_a \psi ~ V^a \\
                 & & df=0 \\
                 & & H= f~V^a V^b V^c \epsilon_{abc}
                 \ena

\noi In the next Section we relate the group manifold formulation of $N=1$, $d=3$ supergravity to its
superspace formulation.

\sect{A bridge between superspace and component actions}
 
We discuss here a technique to relate component to superspace actions, based
on different choices for the Poincar\'e dual that describes the embedding of the spacetime surface inside the supergroup manifold.

As discussed in Section 4, the group manifold action for  a $d$-dimensional supergravity can be written as the superintegral:
\begin{eqnarray}
\label{intA}
S_{SG} = \int_{M^{d|m}} {L}^{d|0} \wedge \eta^{0|m}  \label{superintegral}
\end{eqnarray}
where the Lagrangian ${L}^{d|0}$ is found by using the building rules of Section 5.

Suppose now that 
\eq
dL^{d|0}=0
\en
Then two Poincar\'e duals differing by a total derivative give rise to the same action when inserted into   
(\ref{superintegral}). As a consequence, the action (\ref{superintegral}) can be expressed in multiple ways, using different choices of $\eta$ all in the same cohomology class. This observation can be used to relate component and superspace actions, as we illustrate now in the case of $d=3$ supergravity.
\sk
The Lagrangian $L^{3|0}$  for  $d=3$ supergravity 
 is given in (\ref{d3lagrangian}).  It is a $(3|0)$-form  and, as observed at the end of Section 7.4,
is closed when restricted on fields satisfying the parametrizations (\ref{parRab})-(\ref{pardf}).
Such field configurations are not on-shell since Bianchi identities with parametrizations (\ref{parRab})-(\ref{pardf}) do not imply propagation equations.

 The group manifold action
\eq
S_{3d} = \int_{M^{3|2}} {L}^{3|0} \wedge \eta^{0|2}  \label{3daction}
\en
reproduces the usual component action if we choose $\eta^{0|2}$ to be given by
\eq
  Y^{0|2} = \epsilon_{\al\be} \theta^\al \theta^\be~\epsilon_{\ga\de} \delta (d\theta^\ga) \delta (d\theta^\de) 
  \equiv \theta^2 \delta^2 (d\theta) \label{etacomp}
                 \en
This Poincar\'e dual is closed and not exact, and is an element of the cohomology class $H^{(0|2)}(d, {M}^{3|2})$. The integration over the $d\theta$  and the $\theta$ leads to:
                    \eqa
                   & & S_{d=3} =    \int_{M^3} {L}^{3|0} (\theta=0, d\theta=0) = \nonumber \\
                   & & ~~~~~~ = \int_{M^3} R^{ab} V^c \epsilon_{abc} + 2i \psibar \rho + 6 (f H - {1 \over 2} f^2 V^a V^b V^c \epsilon_{abc})
                     \ena
                      where all forms depend now only on $x$ and have only $dx$ ``legs" because of the two $\theta$'s and $\delta (d\theta)$'s in  $\eta^{0|2}$.  
\sk                                       
Another
Poincar\'e dual can be chosen as follows
\eq
 Y^{0|2}_{ss} =  V^a V^b  (\gamma_{ab})^{\al\be} \iota_\al \iota_\be ~ \delta^2 (\psi)   \label{etass}
  \en
     with                     
   \eq
      \delta^2 (\psi) \equiv  \epsilon_{\ga\de}~ \delta (\psi^\ga) \delta (\psi^\de),~~~  \iota_\al \equiv {\part \over \part \psi^\al},~~~(\ga_{ab})^{\al\be} = (C^{-1})^{\be\ga} (\ga_{ab})^\al_{~\ga} 
     \en
  \noi We use the charge conjugation matrix $C_{\al\be}=\epsi_{\al\be}$ and its inverse $(C^{-1})^{\be\ga} $  to lower and raise spinor indices, with the ``upper left to lower right" convention. We will prove that 
 $Y^{0|2}_{ss}$ is a {\sl bona fide} Poincar\'e dual (closed and not exact) by proving the following
\sk
\noi {\bf Theorem}:  $Y^{0|2}_{ss}$ and $Y^{0|2}$ are in the same cohomology class, i.e.
\eq
Y^{0|2}_{ss} =  Y^{0|2} + d \Omega   \label{3dequivalence}
\en
If the theorem holds, also $Y^{0|2}_{ss}$ is closed and not exact. Moreover, the action 
(\ref{3daction}) computed with $\eta^{0|2}=Y^{0|2}_{ss}$ is equal to the one with $\eta^{0|2}=Y^{0|2}$,
thanks to $dL^{0|3}=0$.
\sk
\noi {\bf Proof:}
\sk
1)  Recall that varying continuously the embedded surface $M^d \subset M^{d|m}$ does not change
the action (\ref{superintegral}) when $dL^{d|0}=0$, since the change in $\eta^{0|m}$ is a total derivative, see (\ref{etachange}). Thus by continuously deforming the soft group manifold to its rigid limit, $Y^{0|2}_{ss}$ gets continuosly connected  to its rigid limit $Y^{0|2}_{rigid}$, obtained from  $Y^{0|2}_{ss}$ by expressing $V$ and $\psi$ with their values on the rigid supergroup manifold $M^{3|2}$. These values are given by the left-invariant vielbein components $V^a$, $\om^{ab}$ and $\psi$: 
  \eqa
  & & V^a = 2dx^a + {i\over 2} \thetabar \ga^a d\theta \\
  & & \om^{ab} =0 \\
  & & \psi = d \theta
  \ena
   and satisfy the Cartan-Maurer equations, i.e. eq.s (\ref{d3cm1})-(\ref{d3cm3}) with
   curvatures = 0. 
 \sk  
   2) Substituting inside (\ref{etass}) yields the rigid Poincar\'e dual
   \eq
   Y^{0|2}_{rigid}=(2dx^a + \thetabar \ga^a d\theta) (2dx^b + \thetabar \ga^b d\theta)
    (\gamma_{ab})^{\al\be} \iota_\al \iota_\be ~ \delta^2 (d\theta) 
    \en
     describing the embedding of flat Minkowski space into the supergroup manifold
$M^{3|2}$. With the help $d=3$ gamma identities (see Appendix C), it is not difficult to show that
\eq
 Y^{0|2}_{rigid} =  \epsilon_{\al\be} \theta^\al \theta^\be~ \delta^2 (d\theta)  +d \Omega' =   Y^{0|2} + d \Omega' 
 \en
  where
  \eq
  \Om'= dx^a~ \thetabar \ga^b \iota~ \thetabar \ga^c \iota ~ \epsi_{abc} \delta^2 (d\theta) + 
  dx^a dx^b (\ga_{ab})^{\al\be} \theta^\ga \iota_\al \iota_\be \iota_\ga \delta^2 (d\theta)
  \en
  up to constant factors. Thus $Y^{0|2}_{rigid}$ is in the same cohomology class of
  $Y^{0|2}$, and because of 1) also in the same cohomology class of 
  $Y^{0|2}_{ss}$, which proves the Theorem $ \Box.$
\sk
Thanks to the above theorem, we  have the equivalence:
                       \eq
                        S_{3d} = \int_{{ M}^{3|2}} { L}^{3|0} \wedge
                        Y^{0|2}= \int_{{ M}^{3|2}} { L}^{3|0} \wedge Y^{0|2}_{ss}
                        \en
                        since $d {L}^{(3|0)}=0$. 
                        
Computing now the action with $Y^{0|2}_{ss}$, we see that only the first two terms of
                        ${L}^{(3|0)}$ contribute, and using
                        the curvature parametrizations for $R^{ab}$ and $\rho$ one finds:
                         \eq
                          S_{3d} = 6i \int_{{M}^{3|2}}  f \epsilon_{abc} V^a V^b V^c \delta^2 (\psi) = 6i \int [d^3x d^2\theta]
                         ~ f ~{\rm Sdet} (E)
                           \en
                           where $E^A=(V^a,\psi^\al)$ is the supervielbein in superspace and we have used
\begin{eqnarray}
\label{volume}
{\rm Vol}^{(3|2)} = \epsilon_{abc} V^a \wedge  V^b \wedge V^c \wedge \delta^2 (\psi) = {\rm Sdet}(E) ~d^3x ~\delta^2(d\theta)
\end{eqnarray}
            Recalling that $f$ is identified with the scalar superfield $R$ we  
            make contact with the superspace action for $d=3$ supergravity.
            The equality (\ref{volume}) can be proven by recalling the formula for the
            superdeterminant of a supermatrix:
            \eq
           Sdet \left(
\begin{array}{cc}
A & B\\
C &D \\
\end{array}
\right) = det(A-BD^{-1}C)(detD)^{-1}
            \en
            applied to the (super)vielbein supermatrix:
            \eq E_\Lambda^A=
            \left(
\begin{array}{cc}
V^a_\mu &  V^a_\be \\
\psi^\al_\mu &  \psi_\be^\al \\
\end{array}
\right),
            \en
The supermatrix $E^A_\Lambda$ is defined by the expansion of $V^a$ and $\psi^\al$ on a coordinate basis:
                        \eqa
               & & V^a=V^a_\mu dx^\mu + V^a_\be d\theta^\be \label{E1} \\
      & &    \psi^\al=\psi^\al_\mu dx^\mu + \psi^\al_\be d\theta^\be \label{E2} 
                    \ena
Substituting (\ref{E2}) into $\de^2 (\psi) \equiv \epsi_{\al \be} \de (\psi^\al) \de (\psi^\be)$ of (\ref{volume})  
produces the identification 
\eq
d\theta^\be=- (\psi^{-1})_\al^\be \psi^\al_\mu dx^\mu
\en
Then the dreibein $V^a$ as expanded in (\ref{E1}) can be written in (\ref{volume}) as
\eq
V^a = (V^a_\mu - V^a_\be (\psi^{-1})^\be_\al \psi^\al_\mu) dx^\mu
\en
and one recognizes the $A-BD^{-1}C$ structure of the $Sdet$. Finally the 
$(det D)^{-1}$ factor in the $Sdet$ arises as the inverse Jacobian $1/det (\psi^\al_\be)$
necessary to express $\epsi_{\al\be} \de (\psi^\al) \de (\psi^\be)$ in terms of 
$\epsi_{\al\be} \de (\theta^\al) \de (\theta^\be)$.
\sk
  In conclusion, the group-manifold Lagrangian ${L}^{(3|0)}$,
                           integrated on superspace, yields
                           both the
                           usual spacetime $d=3$, $N=1$ supergravity action, and its superspace version.

   \sect{Off-shell $N=1, d=4$ supergravity (new minimal)}
  
  \subsection{Off shell degrees of freedom}

The theory contains a vielbein 1-form $V^a$ with 6 off-shell degrees of freedom 
and a Majorana gravitino $\psi^\al$ with 12 off-shell degrees of freedom.
We can match off-shell d.o.f. by adding an auxiliary bosonic 1-form $A$ (3 d.o.f.) and a auxiliary bosonic 2-form $T$ (3 d.o.f.).  The theory with these auxiliary fields was first constructed in ref. \cite{SW}, and recast
in the group manifold formalism in ref. \cite{DFTvN}.

  \subsection{The extended superPoincar\'e algebra}
  
 The starting superalgebra is the superPoincar\'e algebra, extended with a 1-form $A$ and a 2-form $T$.

 The deformed Cartan-Maurer equations for the extended
  soft superPoincar\'e manifold are
   \eqa
   & & R^{ab}=d \omega^{ab} - \omega^a_{~c} ~ \omega^{cb} \\
   & & R^a=dV^a - \omega^a_{~b} ~ V^b -{i \over 2} \psibar \gamma^a \psi \equiv \Dcal V^a -{i \over 2} \psibar \gamma^a \psi\\
   & & \rho = d\psi - {1 \over 4} \omega^{ab} \gamma_{ab} \psi  - {i \over 2} \ga_5 \psi A \equiv \Dcal \psi 
   - {i \over 2} \ga_5 \psi A\\
   & & R^{\square} = dA \\
   & & R^{\otimes}=dT-{i \over 2} \psibar \gamma_a \psi ~V^a
    \ena
    \noi where $\Dcal$ is the Lorentz covariant derivative. These equations can be considered  {\it definitions} for the Lorentz curvature, the (super)torsion, the gravitino field strength and the 1-form and 2-form field strengths respectively.
    The Cartan-Maurer equations are invariant under rescalings 
     \eq
     \omega^{ab} \rightarrow \lambda^0 \omega^{ab}, ~V^a \rightarrow \lambda  V^a,~\psi \rightarrow \lambda^{1\over 2} \psi,~A \rightarrow \lambda^0 A,~T \rightarrow \lambda^2 T \label{rescalings}
      \en           
    \noi Taking exterior derivatives of both sides yields the Bianchi identities:
     \eqa
    & &  \Dcal R^{ab} =0 \\
    & &  \Dcal R^a + R^a_{~b} ~ V^b - i~ \psibar \gamma^a \rho =0\\
    & & \Dcal \rho + {1 \over 2} \gamma_5 \rho A+ {1 \over 4} R^{ab} \gamma_{ab} ~\psi - {i \over 2} \gamma_5 \psi R^\square=0\\
    & & dR^\square=0 \\
    & & dR^{\otimes} - i~ \psibar \gamma_a \rho V^a + {i \over 2} \psibar \gamma_a \psi~ R^a = 0
     \ena
     invariant under the rescalings (\ref{rescalings}).
     
     \subsection{Curvature parametrizations}
     
     According to the rheonomic approach, we parametrize the curvatures so that ``outer" components"
      (i.e. components along at least one fermionic leg) are related to inner components (i.e. components on bosonic legs).
      The most general parametrization compatible with the scalings (\ref{rescalings}) and $SO(3,1) \times U(1)$ gauge invariance 
   is the following:
       \eqa
        & & R^{ab} = R^{ab}_{~~cd} ~V^c V^d + \thetabar^{ab}_{~~c}~\psi ~V^c + i  c_1~\epsilon^{abcd} ~\psibar \gamma_c \psi f_d\\
        & & R^a = 0 \\
        & & \rho = \rho_{ab} V^a V^b + i a \gamma_5 \psi f_a V^a - i c_2  \gamma_5 \gamma_{ab} \psi V^a f^b  \\
        & & R^\square = F_{ab} V^a V^b + \psibar \chi_a V^a + i c_3 \psibar \gamma_a \psi f^a \\
        & & R^{\otimes} = f^a~V^b V^c V^d \epsilon_{abcd} \\
        & & \Dcal f_a= (\Dcal_b f_a)~ V^b + \psibar \Xi_a
        \ena
                The $VV$ component $F_{ab}$ of $F$, and the $VVV$ component $f_a$ of $R^{\otimes}$ scale respectively as $F_{ab} \rightarrow \lambda^{-2} F_{ab}$ and $f_a \rightarrow \lambda^{-1} f_a$.  The Bianchi identities require that: 
         \eq
         c_1=  c_2 = {3 \over 2} ,~c_3 = 3-a
         \en
         \noi and
           \eqa
         & & \thetabar^{ab}_{~~c} = 2 i \rhobar_c^{~[a} \gamma^{b]}  -i ~\rhobar^{ab} \gamma_c\\
         & & \Xi^a= - {i \over 3!} \epsilon^{abcd} \gamma_b \rho_{cd} \\
         & & \chi_a= 2 ( \gamma_5 \gamma^b  \rho_{ab} + {ia \over 3!} \epsilon_{abcd} \gamma^b \rho^{cd})
         \ena

         \noi Note that, thanks to the presence of the auxiliary fields, the Bianchi identities do not imply 
         equations of motion for the spacetime components of the curvatures.
         
         \subsection{The group manifold action}
         
         With the usual group-geometrical methods, the action is determined to be
          \eq
           S_{d=4 SG} = \int_{M^4} R^{ab} V^c V^d \epsilon_{abcd} + 4 \psibar \gamma_5 \gamma_a  \rho V^a - 4 R^\square T +
            \alpha (f_a R^{\otimes} V^a +{1 \over 8} f_e f^e  V^a V^b V^c V^d \epsilon_{abcd})
           \label{spacetimeaction}
            \en
            The action is obtained by taking for the Lagrangian $L^{4|0}$  the most general $SO(3,1) \times U(1)$ 
            scalar 4-form, invariant under the rescalings discussed above, and then 
            requiring that the variational equations admit the vanishing curvatures solution
             \eq
             R^{ab} = R^a = \rho = R^\square=R^{\otimes} = f_a= 0
             \en
                           The remaining parameter $\alpha$ is fixed by requiring the closure of $L^{4|0}$ , i.e. $dL^{4|0} =0$.
               This yields $\alpha = 4(4a-3)$, and ensures off-shell closure of the supersymmetry transformations
               given below. Notice that $a$ is essentially free, since the term $ i a \gamma_5 \psi f_a V^a$ in the 
               parametrization of the gravitino curvature $\rho$ can be reabsorbed into the definition of the 
               $SO(3,1) \times U(1)$-covariant derivative on $\psi$, by redefining $A'=A + 2 a f_a V^a$. 
               Choosing $a={3 \over 4}$ simplifies the action, reducing it to the first three terms, so that
               the 0-forms $f_a$ do not appear.

               \subsection{Field equations}
               
               Varying $\omega^{ab}$, $V^a$, $\psi$, $A$, $T$ and $f$ in the action (\ref{spacetimeaction}) leads to the   
                equations of motion:
                \eqa
                 & & 2 \epsilon_{abcd} R^c V^d =0 ~~\Rightarrow ~R^a=0 \\
                 & & 2 R^{bc} V^d \epsilon_{abcd}-4 \psibar \gamma_5 \gamma_a \rho + \alpha 
                 (-f_a R^\otimes + {1\over 2} f_e f^e \epsilon_{abcd} V^b V^c V^d) = 0 \\
                 & & 8 \gamma_5 \gamma_a \rho V^a - 4 \gamma_5 \gamma_a R^a - i \alpha \gamma_a \psi V^a
                 f_b V^b =0 \\
                 & & R^\otimes = 0 \\
                 & & -4 R^\square + \alpha (V^a \Dcal f_a - {i \over 2} f_a \psibar \gamma^a \psi - f_a R^a) =0\\
                 & & R^\otimes = f^a V^b V^c V^d \epsilon_{abcd}
                 \ena
                 These equations are satisfied by the curvatures parametrized as in Section 9.3
          and also imply:
                  \eq
                  R^a=R^\square = R^\otimes = f_a = 0
                   \en
                   \eqa
                   & & R^{ac}_{~~bc} - {1 \over 2} \delta^a_b R^{cd}_{~~cd} =0 ~~~(Einstein~eq.)\\
                   & & \gamma^a \rho_{ab}=0~~~(Rarita-Schwinger~eq.)
                   \ena
                   The theory has therefore the same dynamical content as the usual $N=1$, $d=4$ supergravity
                   without auxiliary fields.

               \subsection{Off-shell supersymmetry transformations}
               
               Supersymmetry transformations are obtained by applying the Lie derivative along 
               the fermionic directions (i.e. along tangent vectors dual to $\psi$):
                \eqa
              & &  \delta_\epsi V^a = -i \psibar \gamma^a \epsi \\
              & &  \delta_\epsi  \psi = \Dcal \epsi + {i \over 2} \gamma_5 A \epsi + i a \gamma_5 \epsi f_a V^a - 
               {3 i \over 2} \gamma_5 \gamma_{ab} \epsi V^a f^b\\
               & & \delta_\epsi A= {\bar \epsi } ({ia\over 3} \epsilon^{abcd} \gamma_b \rho_{cd} - 2 \gamma_5 \gamma_b \rho^{ba})
               V_a \\
             & & \delta \omega^{ab}= \thetabar^{ab}_{~~c} ~ \epsi V^c + 3i \epsilon^{abcd} \psibar \gamma_c \epsi f_d \\
              & & \delta_\epsi  T =  i \psibar \gamma_a \epsi V^a \\
              & & \delta_\epsi f^a = \epsilonbar ~ \Xi^a
                \ena
               and close on all the fields without need of imposing the field equations.
               
               \subsection{The superspace action}
               
               The action (\ref{spacetimeaction}) originates from a 4-form lagrangian $L^{4|0}$ integrated on a 
               4-dimensional bosonic submanifold of the (soft) group manifold $\Gtilde$ = superPoincar\'e in $d=4$. 
             This group-manifold action
               can be written as an integral on the $M^{4|4}$ superspace:
                \eq
                I = \int_{M^4} L^{4|0}= \int_{M^{4|4}} L^{4|0} \wedge \eta^{0|4}_{M^4} \label{superaction}
                 \en
                 where $\eta^{0|4}_{M^4}$ is the {\it Poincar\'e dual} of the $M^4$ bosonic submanifold embedded into 
                 $M^{4|4}$. To retrieve the usual spacetime action one chooses for the Poincar\'e dual the following $(0|4)$-form:
                 \eq
                    \eta^{0|4}_{M^4}=\theta^4 \delta (d\theta)^4 \label{eta4st}
                 \en
                 with
                 \eq
                 \theta^4= \epsilon_{\al\be\gamma\delta} \theta^\al \theta^\be \theta^\gamma \theta^\delta,~~~
                 \delta (d\theta)^4 = \epsilon_{\al\be\gamma\delta} \delta (d\theta^\al ) \delta (d\theta^\be)
                  \delta ( d\theta^\gamma) \delta ( d\theta^\delta)
                  \en
                              Berezin integration in (\ref{superaction})  yields an ordinary spacetime action, integrated on $M^4$:
                    \eq
                     \int_{M^4} L^{4|0} (\theta=0, d\theta=0)  \label{componentaction}
                     \en
                      where all forms depend only on $x$ because of the 4 $\theta$'s in $\eta_{M^4}$, and have only 
                      $dx$ legs because of the 4 $\delta (d\theta)$'s in $\eta_{M^4}$. 
                      
                      Since the $(4|0)$-form (\ref{eta4st}) is closed and not exact, it is a representative of the de Rahm cohomology class $H^{4|0}$.
                   \sk
                   Also in this case we can relate the component action (\ref{componentaction}) to
                   the superspace action discussed for example in  ref.s \cite{GGRS,RuizPvN,WB}. 
                   Indeed $dL^{0|4} =0$, and the  
                   same mechanism
                   used in $d=3$ supergravity can be exploited. For this we refer to \cite{if6}.

\sect{Gauge supergravities}

We give in this Section a brief account of ``gauge supergravities", i.e. theories where the local supersymmetry
is realized as part of a gauged superalgebra. These theories are gauge invariant under
a supergroup of transformations, so that supersymmetry ``lives" on the fiber, and does not mix
with diffeomorphisms on the base space.  The gauge supersymmetry paradigm has been explored
since long ago \cite{Cham,MDM,Townsend1977,PvN2}. Here we treat separately the odd and even dimensional cases, as they involve different constructive procedures . Indeed all gauge supergravity actions  are written in terms of (products of) connection and curvature of a supergroup $G$, but odd-dimensional actions are Chern-Simons actions invariant under the whole $G$,  while even dimensional actions
are invariant only under a subgroup $F$ of $G$. This subgroup may include also part of the supersymmetries
of $G$, and the resulting theory is then locally supersymmetric.
\sk
Two explicit constructions are given in detail: the $d=5$ Chern-Simons supergravity action \cite{Cham2,Cham3}, and the $d=4$
Mac Dowell-Mansouri action \cite{MDM}. For other odd-dimensional CS supergravity actions we refer to
\cite{Tronc,Zanelli,Zanelli2}, while even-dimensional $d=10+2$ and $d=2+2$ gauge supergravity actions have been constructed in 
\cite{d12} and \cite{d2p2} respectively.

\subsection{Gauge supergravities in odd dimensions}

Chern-Simons (CS) supergravities \cite{Cham2,Cham3,Tronc,Zanelli,Zanelli2} offer interesting alternatives to standard supergravities, since

$\bullet$ ~~supersymmetry is realized as a {\sl gauge} symmetry, part of a gauge supergroup $G$ under which the CS  Lagrangian is invariant up to a total derivative. The superalgebra closes off-shell by construction, without need
of auxiliary fields.

$\bullet$ ~~the gauge supergroup contains the (anti)-De Sitter superalgebra, so that the theory
is translation-invariant and does not have dimensionful coupling constants. Group contraction
can be used to recover the Poincar\'e superalgebra. Retrieving the Einstein-Hilbert term in this limit
is problematic, but there are techniques (S-expansion method \cite{Izaurieta2009}) that allow to recover Poincar\'e gravity
from CS gravity with a particular "expanded" gauge algebra. 

$\bullet$ ~~ CS gravities are also a particular example of Lovelock gravities \cite{Lovelock}, with at most second order field equations for the metric.

$\bullet$ ~~there is no automatic matching between bosonic and fermionic degrees of freedom, at least off-shell.
Indeed the matching results from superPoincar\'e spacetime symmetry, and fields transforming as vector multiplets under supersymmetry. These assumptions do not hold in CS supergravities: the spacetime symmetry is (anti) de Sitter, and the fields are part of a connection belonging to the adjoint representation of a superalgebra.

These features can be relevant for a consistent  quantization of the theory \cite{Zanelli},
and may give arguments for supersymmetry even if phenomenology seems to rule out
the superpartners one expects from Bose-Fermi matching.

CS gravities and supergravities live only in odd dimensions $D=2n-1$, and contain,
besides the usual Einstein-Hilbert term and its supersymmetrization, also a cosmological term (in the uncontracted version)
and  higher powers of the curvature $2$-form $R$ up to order $n-1$. 

\subsubsection{Chern-Simons forms}

We consider the Chern-Simons  ($2n-1$)-forms $L_{CS}^{(2n-1)}$ defined by
\eq
 dL_{CS}^{(2n-1)} = Tr(R^n)  \label{CSdef}
 \en
 where $R^n \equiv R \we R \we \cdots \we R $ ($n$ times),  and $R=d\Om - \Om \we \Om$  is the curvature $2$-form.
The CS form $L_{CS}^{(2n-1)}$ contains (exterior
products of) the $G$ gauge potential one-form $\Om$ and its exterior derivative. The (super)trace $Tr$ is taken on some representation of the (super)group $G$.

  Thus the CS action is related to a topological action in $2n$ dimensions  via Stokes theorem:
 \eq
  \int_{\part M} L_{CS} ^{(2n-1)}= \int_{M} Tr(R^n)  \label{Stokes}
 \en
 Gauge transformations are defined by
 \eq
 \de_\epsi \Om = d \epsi - \Om \epsi + \epsi \Om,~~~ \Rightarrow ~~~\de_\epsi R = - R \epsi + \epsi R \label{gaugetransf}
 \en
 so that $Tr(R^n)$ is manifestly gauge invariant. Therefore also the CS action
is gauge invariant.

  The CS Lagrangian is given in terms of $\Om$ and  $d\Om$ (or $R$) by the following expressions \cite{Nakahara,EGH}:
   \eq
    L_{CS}^{(2n-1)}= n \int^1_0 Tr[\Om(t d\Om-t^2 \Om^2)^{n-1} ] dt= n \int^1_0 t^{n-1} Tr[\Om(R+(1-t)\Om^2)^{n-1} ] dt
     \en
     For example:
      \eqa
      & & L^{(3)}_{CS} = Tr[R\Om + {1 \over 3} \Om^3] \\
      & & L^{(5)}_{CS} = Tr[R^2\Om + \unmezzo R \Om^3 + {1 \over 10} \Om^5] \label{CS5}\\
        & & L^{(7)}_{CS} = Tr[R^3\Om + {2 \over 5} R^2 \Om^3 + {1 \over 5} R \Om^2 R \Om +  {1 \over 5} R \Om^5
        +{1 \over 35} \Om^7]
         \ena
  
\noi Considering $L_{CS}^{(2n-1)}$ as a function of $\Om$ and $R$, a convenient formula for its gauge variation  is \cite{LCCS} :
  \eq
   \de_\epsi L_{CS}^{(2n-1)} = d ( j_\epsi L_{CS}^{(2n-1)}) \label{CSvariation}
   \en
 \noi where $j_\epsi$ is a contraction acting selectively on $\Om$, i.e.
  \eq
   j_\epsi \Om = \epsi,~~~   j_\epsi R = 0
    \en
    with the graded Leibniz rule $j_\epsi (\Om \Om) = j_\epsi (\Om) \Om - \Om  j_\epsi (\Om) = \epsi \Om - \Om \epsi$ etc. 
\sk\sk

\subsubsection{$d=5$ Chern-Simons supergravity}

The relevant supergroup for $d=5$ CS supergravity is $SU(2,2|N)$ (for a group-geometric construction of standard
$D=5$ supergravity see for ex. \cite{gm21}, p. 755). Indeed this supergroup must contain 
the Poincar\'e  or the uncontracted de Sitter group in $d=5$, i.e. $SO(2,4)$. We discuss here
the uncontracted case: the supergroup extension with N supersymmetries is then $SU(2,2|N)$
(recall the local isomorphism $SO(2,4) \approx SU(2,2)$).

We begin by writing
 the connection and curvature supermatrices.
The gauge connection $1$-form is given by:
 \eq
  \Ombold \equiv 
\left(
\begin{array}{cc}
  \Om^\al_{~\be} &  \psi_j^\al   \\
 -\psibar^i_\be &  A^i_{~j}   \\
\end{array}
\right), ~~~ \Om^\al_{~\be} \equiv (\unquarto \om^{ab} \ga_{ab} - {i \over 2} V^a \ga_a + {i \over 4} b I)^\al_{~\be}, ~~~ 
A^i_{~j} ={i \over N} b \de^i_j + a^i_{~j} 
    \label{Omdef}  
  \en
  where the bosonic $U(2,2)$ subgroup is gauged by the $1$-forms $\om^{ab}$ (spin connection), $V^a$ (vielbein) and $b$ ($U(1)$ gauge field); the antihermitian matrix-valued $1$-forms  $a^i_{~j}$ ($i,j=1...N$) gauge the $SU(N)$ bosonic subgroup; finally the $N$ gravitino $1$-form fields $ \psi_j$ gauge the $N$ supersymmetries. The Dirac conjugate is defined as $\psibar = \psi^\dagger \ga_0$. 
  
The corresponding curvature supermatrix $2$-form is
 \eq
      \Rbold =  d \Ombold - \Ombold \we \Ombold~
  \equiv  \left(
\begin{array}{cc}
   R + \psi_i \we \psibar^i &  \Sigma_j    \\
 -\Sigmabar^i  &  F^i_{~j}+ \psibar^i \we \psi_j   \\
\end{array}
\right) \label{Rdef}
       \en
\noi with\footnote{we omit wedge products between forms}
   \eqa
    & & R = d\Om - \Om  \Om \equiv \unquarto R^{ab} \ga_{ab} - {i \over 2} R^a \ga_a +{i\over 4} r I \label{defR}\\
    & & \Sigma_j = d \psi_j - \Om   \psi_j - \psi_k   A^k_{~j} \equiv D \psi_j \label{defSigma} \\
    & & \Sigmabar^i = d \psibar^i - \psibar^i  \Om  - A^i_{~k}   \psibar^k \equiv D \psibar^i \\
    & & F^i_{~j}= dA^i_{~j} - A^i_{~k}  A^k_{~j}
    \ena
    \noi Immediate algebra yields the components of the $U(2,2)$ curvature $R$:
   \eqa
     & & R^{ab} = d \om^{ab} - \unmezzo \om_c^{~[a}  \om^{b]c}  + \unmezzo V^{[a}  V^{b]}  \\
     & & R^{a} = d V^{a} - \om^a_{~b} ~ V^{b}  \\
    & & r~~ = db
        \ena

 \noi A direct consequence of the curvature definition (\ref{Rdef}) is the Bianchi identity
  \eq
   d\Rbold = - \Rbold  \Ombold + \Ombold   \Rbold
    \en
  \noi which becomes, on the supermatrix entries 
   \eqa
    & & dR = - R \Om+\Om  R,~~dF=-F A + A  F, \label{BianchiR}\\
    & & d\Sigma= -R  \psi + \Om  \Sigma - \Sigma   A + \psi   F, \label{BianchiSigma}\\
    & & d\Sigmabar= - \Sigmabar  \Om + \psibar  R - F  \psibar + A   \Sigmabar
     \label{BianchiSigmabar}\ena
     \sk
     \sk
\noi {\bf $SU(2,2|N)$ gauge transformations}
\sk

\noi The gauge transformations  (\ref{gaugetransf})  close on the Lie (super)algebra:
  \eq
    [\de_{\epsibold_1},  \de_{\epsibold_2}] = \de_{\epsibold_1 \epsibold_2 - \epsibold_2  \epsibold_1}
     \en
 
\noi In the case at hand the $SU(2,2|N)$  gauge parameter is
given by the supermatrix
 \eq
  \epsibold \equiv 
\left(
\begin{array}{cc}
  \epsi^\al_{~\be} &  \epsilon_j^\al   \\
 -\epsilonbar^i_\be &  \eta^i_{~j}   \\
\end{array}
\right), ~~~ \epsi^\al_{~\be} \equiv (\unquarto \epsi^{ab} \ga_{ab} - {i \over 2} \epsi^a \ga_a + {i \over 4} \epsi I)^\al_{~\be}, ~~~ 
\eta^i_{~j} ={i \over N} \epsi \de^i_j + \epsi^i_{~j} 
\en
\sk
\noi and the gauge variations (\ref{gaugetransf}) on the block entries of $\Om$ read
\eqa
& &   \de \Om = d \epsi - \Om  \epsi + \epsi  \Om + \psi_i \epsilonbar^i + \epsilon_i \psibar^i \\
& & \de \psi_i = d \epsilon_i - \Om  \epsilon_i + \epsilon_j  A^j_{~i} - \psi_j  \eta^j_{~i} + \epsi  \psi_i 
\label{psivariation} \\
& & \de \psibar^i = d \epsilonbar^i +\epsilonbar^i  \Om - A^i_{~j}  \epsilonbar^j +  \eta^i_{~j}  \psibar^j  - \psibar^i  \epsi   \\
 & & \de A^i_{~j} = d  \eta^i_{~j} - A^i_{~k}   \eta^k_{~j} +  \eta^i_{~k}  A^k_{~j} + \psibar^i  \epsilonbar_j - \epsilonbar^i  \psi_j
\ena
\noi On the $\Om$ component fields they take the form
 \eqa
  & & \de \om^{ab} = d \epsi^{ab} -\om_c^{~[a}  \epsi^{b]c} + \epsi_c^{~[a}  \om^{b]c} + 2 V^{[a}  \epsi^{b]} +  \unmezzo (\psibar  \ga^{ab} \epsi - \epsilonbar  \ga^{ab} \psi) \\
   & & \de V^a = d \epsi^a -\om^{ab}  \epsi^b  +
    V^b  \epsi^{ab}  - i(\psibar  \ga^a \epsilon - \epsilonbar  \ga^a \psi)  \\
     & & \de b = d \epsi -  i(\psibar  \epsilon - \epsilonbar  \psi) 
   \ena
For $N=4$  the supergroup $SU(2,2|N)$ is not simple
anymore and the $U(1)$ gauged by the $b$ field becomes a central extension. Consider now the
$U(1)$ gauge variation of the gravitini, cf. (\ref{psivariation}):
 \eq
 \de \psi_i =  i ~({1 \over 4} - {1\over N} )~\epsi \psi_i 
 \en
\noi For $N=4$ we see that the gravitini become uncharged with respect to this $U(1)$.
\sk\sk
\noi {\bf The action}
\sk
  
Substituting $\Rbold$ and $\Ombold$ into (\ref{CS5}), we obtain
the $d=5$ CS action invariant under the $SU(2,2|N)$ gauge variations of the preceding subsection. 
The result is
  \eq
  \int Str( L^{(5)}_{CS}) = \int L_{U(2,2)} + L_{A}+ L_{fermi} \label{starCSSG5}
    \en
 \noi with
 \eqa
 & &  L_{U(2,2)} =  Tr[ R  R  \Om + \unmezzo R  \Om^{ 3} + {1 \over 10} \Om^{ 5}] \\
 & & L_A ~~~~=  - Tr[F  F  A + \unmezzo F  A^{ 3} + {1 \over 10} A^{ 5} ]\\
& &  L_{fermi} = {3 \over 2}  \psibar  (R  \Sigma + \Sigma  F) + {3 \over 2} \Sigmabar  (R  \psi + 
\psi  F)  \nonumber \\
 & & ~~~~~~~~~~+  \psibar  \psi  (\psibar  \Sigma + \Sigmabar  \psi ) 
 \ena
This is
the action discussed in refs. \cite{Cham2,Tronc,Zanelli}. The $b$ field kinetic term has two contributions, from the $R R  \Om$ and the $F  F  A$ terms, and is proportional to:
 \eq
  ({1 \over 16} - {1 \over N^2}) (db ~ db ~ b)
 \en
\noi vanishing for $N=4$. 
\sk
We can obtain a slightly more explicit form for
$\int  L_{U(2,2)} $ by splitting the $U(2,2)$ connection in its ``Lorentz + rest" parts as 
 \eq
 \Om =  \om + V, ~~~\om \equiv \unquarto \om^{ab} \ga_{ab},~~~V \equiv  - {i \over 2} V^a \ga_a + {i \over 4} b I
 \en
 and correspondingly the $U(2,2)$ curvature as
 \eq
 R = \Rcal + T - V  V, ~~~\Rcal \equiv d\om - \om  \om, ~~~T \equiv dV - \om  V - V  \om
 \en
 Then we find, after some integrations by parts and use of the Bianchi identities (\ref{BianchiR})-(\ref{BianchiSigmabar}):
 \eqa
 & &  \int L_{U(2,2)} = 3 \int Tr[  \Rcal  \Rcal  V - {2\over 3} \Rcal  V^{ 3} + {1 \over 5} V^{ 5} \nonumber \\
  & & ~~~~~~~~~~~+ \unmezzo (T  \Rcal + \Rcal  T)  V + {1 \over 3} T  T  V - \unmezzo T  V^{ 3}] \nonumber \\
& & ~~~~~~~~~~~~ + \int Tr[\Rcal  \Rcal  \om + \unmezzo \Rcal  \om^{ 3} + {1 \over 10} \om^{ 5}] \label{LU22}
  \ena
\noi The last line is the integral of the Lorentz CS form $L_{Lor}$. Its derivative gives the Pontryagin $6$-form:
  \eq
d L_{Lorentz} = Tr[ \Rcal  \Rcal  \Rcal]
  \en
This $6$-form $Tr[ \Rcal  \Rcal  \Rcal]$ vanishes identically, so that the last line in 
({\ref{LU22}) can be deleted by virtue of (\ref{Stokes}).

\subsection{Gauge supergravities in even dimensions}

\subsubsection{The $d=4$ Mac Dowell-Mansouri action }

This Section follows closely ref. \cite{Castellani2013}. The Mac Dowell-Mansouri action \cite{MDM} is a $R^2$-type reformulation of (anti)de Sitter supergravity in $D=4$.
It is based on the supergroup $OSp(1|4)$, and the fields $V^a$ (vierbein), $\omega^{ab}$ (spin connection) 
and $\psi$ (Majorana gravitino) are 1-forms contained in the  $OSp(1|4)$ connection $\Ombold$, in a $5 \times 5$ supermatrix representation:

\eq
  \Ombold \equiv 
\left(
\begin{array}{cc}
  \Om &  \psi    \\
 \psibar  &  0   \\
\end{array}
\right), ~~~ \Om \equiv \unquarto \om^{ab} \ga_{ab} - {i \over 2} V^a \ga_a
    \label{Omdef4}  
  \en
The corresponding $OSp(1|4)$ curvature supermatrix is
 \eq
      \Rbold =  d \Ombold - \Ombold \we \Ombold~
  \equiv  \left(
\begin{array}{cc}
   R &  \Sigma    \\
 \Sigmabar  &  0   \\
\end{array}
\right) \label{Rdef4}
       \en
       \noindent and straightforward matrix algebra yields:
   \eqa
    & & R = \unquarto R^{ab} \ga_{ab} - {i \over 2} R^a \ga_a  \label{defR4}\\
    & & \Sigma = d \psi - \unquarto \om^{ab} \ga_{ab} \psi + {i \over 2} V^a \ga_a \psi \label{defSigma4} \\
    & & \Sigmabar = d \psibar - \unquarto \psibar ~\om^{ab} \ga_{ab} + {i \over 2}  \psibar V^a \ga_a \\
        & & R^{ab} \equiv d \om^{ab} - \om^{a}_{~c} ~\om^{cb} + V^a V^b + {1 \over 2} \psibar \ga^{ab} \psi \\
     & & R^{a} \equiv d V^{a} - \om^{a}_{~b} V^{b}  -  {i \over 2} \psibar \ga^a \psi 
     \ena
 \noi We have also used the Fierz identity for $1$-form Majorana spinors:
  \eq
  \psi \psibar = \unquarto ( \psibar \ga^a \psi \ga_a - \unmezzo \psibar \ga^{ab} \psi \ga_{ab} )
  \en
\noi (to prove it, just multiply both sides by $\ga_c$ or $\ga_{cd}$ and take the  trace on spinor indices).  

The Mac Dowell-Mansouri action can be written in terms of the $OSp(1|4)$ curvature $\Rbold$ as:
 \eq
     S =  4 \int STr ( \Rbold \Gbold \Rbold \Gammabold)    \label{MDMaction}
    \en
    where $STr$ is the supertrace and $\Gbold, \Gammabold$ are the following constant matrices:
     \eq
       \Gammabold \equiv  \left(
\begin{array}{cc}
  i \ga_5 &  0    \\
 0 &  0   \\
\end{array}
\right), ~~~ \Gbold = \onebold + {\Gammabold^2 \over 2}= {1 \over 2} \left(
\begin{array}{cc}
 1 &  0    \\
 0 &  2   \\
\end{array}
\right) \label{Gammabold4}
       \en
\noi All boldface quantities are  5 $\times$ 5 supermatrices. 
Carrying out the supertrace, and then the spinor trace, leads to the familiar expression of the
MacDowell-Mansouri action:
\eq
 S = 2i \int Tr( R \we R \ga_5 + 2 \Sigma \we \Sigmabar \ga_5) = 2 \int \unquarto R^{ab} \we R^{cd} \epsi_{abcd} - 2i \Sigmabar \we \ga_5 \Sigma \label{MDMaction2}
  \en
  After inserting the curvature definitions the action takes the form
   \eq
    S = \int \Rcal^{ab} V^c V^d \epsi_{abcd} + 4 \rhobar \ga_a \ga_5 \psi V^a + \unmezzo (V^a V^b V^c V^d + 2 \psibar \ga^{ab} \psi V^c V^d ) \epsilon_{abcd}  \label{adsSG}
    \en
  with
   \eq
    \Rcal^{ab} \equiv d \om^{ab} - \om^{a}_{~c} ~\om^{cb} , ~~\rho \equiv d \psi - \unquarto \om^{ab} \ga_{ab} \psi \equiv \Dcal \psi
     \label{RLordef}
     \en
   We have dropped the topological term $\Rcal^{ab} \Rcal^{cd} \epsilon_{abcd}$ (Euler form), and used 
   the gravitino Bianchi identity 
    \eq
     \Dcal \rho = - \unquarto \Rcal^{ab} \ga_{ab} \psi
   \en
   and the gamma matrix identity $2 \ga_{ab} \ga_5 = i \epsilon_{abcd} \ga^{cd}$  to recognize that $\unmezzo \Rcal^{ab} \psibar \ga^{cd} \psi \epsilon_{abcd} - 4 i \rhobar \ga_5 \rho$
    is a total derivative. The action (\ref{adsSG}) describes $N=1$, $D=4$ anti-De Sitter supergravity,
    the last term being the supersymmetric cosmological term. After rescaling the vielbein and the gravitino as
    $V^a \rightarrow \lambda V^a$, $\psi  \rightarrow \sqrt{\lambda} \psi$ and dividing the action by $\lambda^2$, the usual (Minkowski)
    $N=1$, $D=4$ supergravity is retrieved by taking the limit $\lambda \rightarrow 0$. This corresponds to the
Inon\"u-Wigner contraction of $OSp(1|4)$ to the superPoincar\'e group.
\sk
\noi {\bf Invariances}
\sk
 \noi As is well known, the action (\ref{MDMaction}), although a bilinear in the $OSp(1|4)$ curvature, is {\it not} invariant under the 
  $OSp(1|4)$ gauge transformations:
   \eq
   \de_{\epsibold} \Ombold = d \epsibold - \Ombold \epsibold + \epsibold \Ombold~ \Longrightarrow~  \de_{\epsibold} \Rbold =  - \Rbold \epsibold + \epsibold \Rbold   \label{Rgauge4}
       \en
       where $\epsibold$ is the $OSp(1|4)$ gauge parameter:
        \eq
         \epsibold \equiv 
         \left(
\begin{array}{cc}
 \unquarto  \epsi^{ab} \ga_{ab} - {i \over 2} \epsi^a \ga_a &  \epsilon   \\
 \epsilonbar  &  0   \\
\end{array}
\right)   \label{gaugeparam4}
 \en
 In fact it is not a Yang-Mills action (involving the exterior product of $\Rbold$ with its Hodge dual), nor a topological action  $\int STr(\Rbold \Rbold)$: the constant supermatrices $\Gbold$ and  $\Gammabold$ ruin the $OSp(1|4)$ gauge invariance, and break it to its Lorentz subgroup.  Indeed the gauge variation of the action (\ref{MDMaction})
 \eq
\delta S = 4 ~ \int STr (\Rbold [\Gbold,\epsibold] \Rbold \Gammabold + \Rbold \Gbold \Rbold [\Gammabold,\epsibold] ) \label{gaugevariation4}
 \en
  vanishes when $\epsibold$ commutes with $\Gammabold$ (and therefore with $\Gbold$), and this happens
  only when $\epsibold$ in (\ref{gaugeparam4}) has $\epsi^{a} = \epsilon =0$, so that only Lorentz rotations leave the action invariant. 
  
  Specializing the gauge parameter $\epsibold$ to describe supersymmetry variations (i.e. only $\epsilon \not= 0$ in (\ref{gaugeparam4})),
  eq. (\ref{gaugevariation4}) yields the supersymmetry variation of the Mac Dowell-Mansouri action: 
  \eqa
  & & \delta_{susy}S= 2i ~\int (\epsilonbar [\gamma_5 , R] \Sigma + \Sigmabar [ \gamma_5 , R ] \epsilon) \label{susyvariation} \\
  & & ~~~~~~~~ =  -4 \int R^a \Sigmabar \ga_a \ga_5 \epsilon
  \ena
 with $R$ defined in (\ref{defR4}). This variation is proportional to the torsion $R^a$, since only
 $R^a \gamma_a$ in $R$ has a nonzero commutator with $\gamma_5$.
 \
  Therefore in second-order 
 formalism, i.e. using the torsion constraint $R^a=0$ to express $\omega^{ab}$ in terms of $V^a$ and $\psi$,  the action is indeed supersymmetric. Another way to recover supersymmetry is by modifying the
 supersymmetry variation of the spin connection, see for ex. \cite{PvN2}. In both cases supersymmetry is
 not part of a gauge superalgebra: off-shell closure of the supersymmetry transformations is not
 automatic, and indeed necessitates the introduction of auxiliary fields.

\subsubsection{Gauge supergravity in $d=10+2$}

Here we give a short description of $d=10+2$ gauge supergravity, summarizing the results of ref. \cite{d12}.

Supergravity theories in dimensions greater than $d=11$ are believed to be inconsistent, since their reduction
to $d=4$ would produce more than $N=8$ supersymmetries, involving multiplets with spin $\ge 2$, and it is known that coupling of gravity with a finite number of higher spins is problematic. 

On the other hand a twelve dimensional theory with signature (10,2) avoids this difficulty, since fermions can be
both Majorana and Weyl in $d=10+2$, with 32 real components, and therefore giving rise to at most
eight supercharges when reduced to $d=4$. This fact has encouraged over the years various
attempts and proposals (\cite{Castellani1982} - \cite{Hewson1999}) for a twelve-dimensional field theory of supergravity.

A $d=10+2$ structure emerges also from string/brane theory, and has been named $F$-theory \cite{Vafa1996}.
The $OSp(1|32)$ superalgebra, a natural choice for the gauge algebra of a $d=10+2$ supergravity,
is also called $F$ algebra \cite{Hewson1999}.

 \sk
In $d=10+2$ dimensions we can write a geometrical $\Rbold^6$-type action that resembles
 the $\Rbold^2$-type $d=4$ Mac Dowell-Mansouri action:
 \eq
 S_{d=10+2 SG} = \int STr (\Rbold^6 \Gammabold) \label{d12action}
 \en
  where $\Rbold$ is now the 
$OSp(1|64)$ curvature supermatrix two-form, and $\Gammabold$ is a constant supermatrix involving $\ga_{13}$ and breaking $OSp(1|64)$ to a $\Ftilde$ subalgebra that includes the $F$ algebra (see below). Contrary to the $d=4$ case, $N=1$ supersymmetry (with a Majorana-Weyl  supercharge) survives as {\it part of this subalgebra}, and closes off-shell. 

The ``would be" gauge fields of $OSp(1|64)$ are one-forms 
$B^{(n)}$ with $n$=1,2,5,6,9,10 antisymmetric Lorentz indices and a Majorana gravitino $\psi$.
The vielbein and the spin connection are identified with $B^{(1)}$ and  $B^{(2)}$ respectively.
These one-forms are organized into an $OSp(1|64)$ connection, in an explicit 65 $\times$ 65
dimensional supermatrix representation. The constant matrix $\Gammabold$ in (\ref{d12action}) ensures that the action is not topological (similarly to the MDM action) and breaks
$OSp(1|64)$ to a subalgebra $\Ftilde = OSp(1|32) \oplus Sp(32)$, under which the action is invariant\footnote{The $\Ftilde$ algebra contains the $F$ algebra: in fact the $F$ algebra is the $OSp(1|32)$ part 
of $\Ftilde$.}.
 Here part of the supersymmetry of
$OSp(1|64)$ survives, in contrast to the $D=4$ case. Supersymmetry is then a gauge symmetry, and 
closes off-shell. Twelve dimensional Lorentz symmetry $SO(10,2)$ is also part of the $\Ftilde$ gauge symmetry,
so that the action is $SO(10,2)$ invariant.

  Under the action of $\Ftilde$, the  $OSp(1|64)$ fields split into a gauge multiplet and a matter multiplet.
The gauge multiplet contains the $\Ftilde$ gauge fields: the spin connection $B^{(2)}$, a Majorana-Weyl gravitino $\psi_+$, and the other ``even" one-form fields
   $B^{(6)}$ and $B^{(10)}$. The matter multiplet contains the remaining $OSp(1|64)$ fields: the ``odd" one-form fields $B^{(1)}$ (the vielbein), $B^{(5)}$ and $B^{(9)}$, and a Majorana anti-Weyl gravitino $\psi_-$.

\sect{Form hamiltonian}

In this final Section we recall a hamiltonian formalism well-adapted to geometrical theories described by
$d$-form Lagrangians. It goes under the name of ``covariant canonical formalism" (CCF), and has
been first proposed in ref.s \cite{CCF1}-\cite{CCF5}. Recent developments can be found in \cite{CD}.

In fact this formalism is suggested by the form version of the Euler-Lagrange equations (\ref{ELequations1}), discussed in Section 2. Considering the Lagrangian $d$-form as depending on 1-form fields $\phi$ and
2-form ``velocities" $d \phi$ naturally leads to the definition of a $(d-2)$-form momentum:
\eq
\pi \equiv {\partial L \over \partial (d\phi)}
\en
and a $d$-form Hamiltonian density
\eq
H \equiv \pi \we d\phi - L
\en
This Hamiltonian density does not depend on the ``velocities" $d \phi$ since
\eq
{\partial H \over \partial (d\phi)}  = \pi - {\partial L \over \partial (d\phi)}= 0 
\en
Thus $H$ depends on $\phi$ and $\pi$:
\eq
H=H(\phi,\pi)
\en
and the form-analogue of the Hamilton equations reads:
\eq
d \phi = {\partial H \over \partial \pi} ,~~~d \pi =  + {\partial H \over \partial \phi} \label{formHE}
\en
The first equation is equivalent to the momentum definition, and the second is equivalent to the
Euler-Lagrange form equations (note the + sign due to the + sign in (\ref{ELequations1})).
\sk
\noi {\bf Note 1:} the derivative of a $p$-form $F$ with respect to a basic 1-form $\phi$ or 
momentum $d-2$ form $\pi$ is always defined by first bringing $\phi$ or $\pi$ to the left in $F$
(taking into account the sign changes due to the gradings)
and then canceling it against the derivative. In other words, we use the graded
Leibniz rule, considering ${\partial \over \partial \phi}$ to have the same grading as $\phi$, and 
similar for $\pi$.
\sk
As an easy exercise, let us apply the formalism to pure $d=4$ gravity. The fields $\phi$ in this case
are the vierbein $V^a$ and the spin connection $\om^{ab}$. From:
\eq
L (\phi, d\phi) = R^{ab} V^c V^d \epsi_{abcd} = d \om^{ab} V^c V^d \epsi_{abcd} - \om^a_{~e} \om^{eb} V^c V^d \epsi_{abcd}
\en
we find the momenta:
\eqa
& & \pi_{a} = {\partial L \over \partial (dV^a)} = 0 \\
& & \pi_{ab} = {\partial L \over \partial (d \om^{ab})} =V^c V^d \epsi_{abcd}
\ena
and the Hamiltonian density:
\eq
H= dV^a \pi_a + d \om^{ab} \pi_{ab} - d \om^{ab} V^c V^d \epsi_{abcd} + \om^a_{~e} \om^{eb} V^c V^d \epsi_{abcd}
\en
Both momenta definitions are primary constraints:
\eq
\Phi_a \equiv \pi_a = 0,~~~\Phi_{ab} \equiv \pi_{ab} - V^cV^d \epsi_{abcd} = 0
\en
 since they do not involve the ``velocities" $dV^a$ and
$d\om^{ab}$. Therefore $dV^a$ and
$d\om^{ab}$ are undetermined at this stage. Requiring the ``conservation" of
$\Phi_a$ and $\Phi_{ab}$, i.e. their {\it closure} in the present formalism, leads to the
secondary constraints:
\eqa
& & d \Phi_c = 0 ~~~\Rightarrow ~~~ R^{ab} V^d \epsi_{abcd}  = 0 \label{secondary1} \\
& & d \Phi_{ab} = 0 ~~\Rightarrow ~~~  R^c V^d \epsi_{abcd} = 0 \label{secondary2}
\ena
after use of the 
Hamilton equations
\eq
d \pi_c = {\partial H \over \partial V^c},~~~d\pi_{ab} =   {\partial H \over \partial \om^{ab}}
\en
and the
definitions of the curvatures 
\eq
R^a = dV^a -\om^a_{~b} V^b,~~~R^{ab}= d \om^{ab} - \om^a_{~e} \om^{eb} 
\en
To derive (\ref{secondary2}) we also made use of the identity
\eq
F^e_{~[a} \epsi_{bcd]e} =0
\en
holding for any antisymmetric $F$. 
Thus the secondary constraints reproduce the field equations (\ref{PoincarefieldeqRa}), (\ref{PoincarefieldeqRab}) of vierbein gravity. 

No tertiary constraints arise since the secondary constraints (\ref{secondary1}),
(\ref{secondary2}), i.e. the l.h.s. of the field equations, are ``conserved". This can be checked
by applying the exterior differential to the constraints, and using the Bianchi identities
(\ref{bianchiRa}), (\ref{bianchiRab}).
\sk
\noi {\bf Form brackets}
\sk
The differential of any $f$-form $F$ depending on the 1-form fields $\phi$ and their conjugated $(d-2)$ -form momenta $\pi$ can be expressed as
\eq
d F(\phi,\pi)=d \phi {\partial F \over \partial \phi} + d \pi {\partial F \over \partial \pi} =
 {\partial H \over \partial \pi} {\partial F \over \partial \phi} +  {\partial H \over \partial \phi} {\partial F \over \partial \pi}  \label{dF1}
\en
where the graded Leibniz rule for the differential has been taken care of by the
definition of partial derivative given in Note 1 after eq.s (\ref{formHE}), and we have
used the Hamilton equations of motion.

This would suggest a {\it form} analogue of the Poisson bracket of the $f$-form $F$ with
the $d$-form $H$:
\eq
\{ F,H \} \equiv  {\partial H \over \partial \phi} {\partial F \over \partial \pi} +
  {\partial H \over \partial \pi} {\partial F \over \partial \phi} \label{formbracket}
  \en
  so that we recover the familiar looking formula:
  \eq
dF = \{ F,H \} \label{dF2}
\en
To make the definition (\ref{formbracket}) consistent for any $a$-form $A$ and $b$-form $B$
  (with any degree $a \le d$, $b \le d$)  it must be generalized to \cite{CD}:
\eq
\{ A, B \} \equiv  {\dleft B \over \partial \pi^i}~
{\dright A\over \partial \phi_i} -  (-)^{p_i d} ~{\dleft B \over \partial \phi_i} {\dright A\over \partial \pi^i}  \label{FPB}
\en
where $\dright$ indicates the derivative ``acting from the left" as defined in the Note 1 after eq.s (\ref{formHE}), and $\dleft$ is defined in a specular way as ``acting from the right".  A sum on the fields (labelled by the index $i$) is understood, and $p_i$ is the degree
of the form $\phi_i$. Thus the definition (\ref{FPB}) holds also in the case of fundamental fields $\phi_i$ being forms
of arbitrary degree $p_i$, and the covariant hamiltonian formalism can be applied to the
free differential algebras of Section 6.  By using the relations 
\eq
{\dleft F \over \partial \phi_i} = (-)^{p_i (f+1)} ~{\dright F \over \partial \phi_i},  ~~~{\dleft F \over \partial \pi^i} = (-)^{(d-p_i-1) (f+1)} ~{\dright F \over \partial \pi^i}\label{leftright}
\en
one can verify that (\ref{FPB}) indeed reduces to (\ref{formbracket}) for $F$ and $H$.
\sk
\noi {\bf Note 2:} The form Poisson bracket between the $a$-form $A$ and the $b$-form $B$ is a ($a+b-d+1$)-form, and canonically conjugated forms satisy:
\eq
 \{ \phi_i, \pi^j\}  = \delta_i^j \label{canonicalPB}
 \en

\noi Using the definition (\ref{FPB}), the following relations can be shown to hold:
\eqa
& & \{ B,A \} = - (-)^{(a+d+1)(b+d+1)} \{ A,B \}  \label{prop1} \\
& & \{A,BC \} = B \{A,C \} + (-)^{c(a+d+1)} \{A,B \} C \\
& & \{AB,C \} =  \{A,C \} B + (-)^{a(c+d+1)}  A \{B,C \}  \\
& & (-)^{(a+d+1)(c+d+1)} \{ A, \{ B,C \} \} + cyclic~=0\\
& & (-)^{(a+d+1)(b+d+1)} \{  \{ B,C \},A \} + cyclic~=0 \label{prop5}
\ena
i.e. graded antisymmetry, derivation property, and form-Jacobi identities.
\sk
\noi {\bf Note 3:} a different definition of form Poisson bracket was given in ref. \cite{CCF1}, based on postulated properties of the FPB
rather than on the Legendre transformation that leads to the evolution equation (\ref{dF2}). In fact the properties of the FPB in 
\cite{CCF1} differ from the ones given above, which are {\it deduced} from the definition (\ref{FPB}). 
\sk

  Using the form bracket we find the constraint algebra:
  \eq
  (\Phi_a,\Phi_b)=  (\Phi_{ab},\Phi_{cd})=0;~~~  (\Phi_a,\Phi_{bc})=-2\epsi_{abcd} V^d
  \en
 showing that the constraints are not all first-class. There is, however, a first-class combination of the constraints:
  \eq
  R^{ab} \Phi_{ab} + R^a \Phi_a
  \en
  
  One may continue the constraint analysis, separating first-class from second-class constraints,
  constructing the form analogue of the Dirac bracket, etc. In part this has been done in ref. \cite{CCF2}, where the correspondence with the ``usual" hamiltonian formalism
  for first order tetrad gravity of ref. \cite{CVF} was established, and extended to
  canonical supergravity \cite{CCF3}. It could be worthwhile to recast in
  covariant hamiltonian language the algorithm for the construction of gauge generators of ref. \cite{SCHS}.
  For $d=3$ and $d=4$ gravity this has been done in ref. \cite{CD}.

\section*{Acknowledgements}

It is a pleasure to acknowledge useful discussions with my colleagues and friends Laura Andrianopoli,  Paolo Aschieri, Roberto Catenacci, Anna Ceresole, Alessandro D'Adda, Riccardo D'Auria, Pietro Fr\'e,  Ferdinando Gliozzi, Pietro Antonio Grassi,  Massimo Porrati, Mario Trigiante and Jorge Zanelli.

\appendix

 \sect{Group manifold geometry}

This brief resum\'e is taken from Sec. 2 of \cite{gm23}.
We start from a Lie algebra Lie({\sl G}), with generators $T_A$ 
satisfying the commutation relations
\eq
[T_A,T_B]={C^C}_{AB}T_C  \label{2.1}
\en
\noi For simplicity we consider only usual Lie algebras. The extension to superalgebras
is straightforward and only necessitates extra phases (for ex. anticommutators
for fermionic generators) due to gradings. 

A generic group element $g \in G$ connected with the identity
\footnote{Hereafter $G$ indicates the part of the group connected 
with the identity.} can be expressed as
\eq
g=\exp (y^A T_A) \equiv y \label{2.2}
\en
\noi where $y^A$ are the (exponential) coordinates of the group 
manifold. Each element of $G$ is labelled by the coordinates $y^A$, and 
for notational economy we denote it simply by $y$. Similarly $yx$ stands 
for ~$\exp (y^A T_A) \exp (x^B T_B)$, the product of two group elements, 
and by $(yx)^M$ we denote the corresponding coordinates.

Consider now $(yx)^M$ as a function\footnote{Since $G$ is a Lie 
group, this function is smooth.} of $x^A$:
\eq
(yx)^M=y^M+e_A^{~~M} (y) x^A + e_{AB}^{~~~M} (y) x^A x^B+ ... \label{2.3}
\en
\noi For infinitesimal $x$:
\eq
(yx)^M=y^M+(x^A t_A) y^M= (1+x^A t_A) y^M, ~~~t_A 
\equiv e_A^{~~N} (y) \pdyN \label{2.4}
\en
\noi so that the $t_A$ are a differential representation of the 
abstract generators $T_A$, and satisfy therefore the same algebra:
\eq
[t_A, t_B]=\CCAB t_C \label{2.5}
\en
The geometrical meaning of the components $e_A^{~N} (y)$ in eq. (\ref{2.3}) 
is clear: consider the infinitesimal displacement $\de_A y^M$ due to the 
(right) action of $1+ \varepsilon T_A$ ($\epsi$ = infinitesimal 
parameter). Then 
\eq
\de_A y^M = \epsi e_A^{~~M} (y) \label{2.6}
\en
\noi and the dim$G$ vectors $e_A^{~~M} (y)$, A=1,...dim$G$ are simply 
the tangent vectors at $y$ in the direction of the displacements $\de_A 
y^M$. It is customary to call tangent vector along the $T_A$ 
direction the whole differential operator $t_A \equiv 
e_A^{~~N} (y) \pdyN$. 

Note that $e_A^{~~M}$ is an invertible matrix, 
since the map $y \rightarrow yx$ is a diffeomorphism.

The $t_A (y)$ span the tangent space of $G$ at $y$: they form a 
contravariant basis. The ``coordinate" 
basis given by the vectors $\pdyN$
is related to the $t_A$ (the intrinsic basis) 
via the nondegenerate matrix $e_A^{~~N}$.
The indices A,B,... are tangent space indices (``flat" indices) and are
inert under $y$ coordinate transformations. The indices M,N,... are
coordinate indices (``world" indices) and do transform under coordinate
transformations in the usual way (see later).   
Next we define the one-forms $\si^A (y)$ as the duals of the $t_A$:
\eq
\si^A (t_B) = \de_A^B \label{2.8}
\en
\noi The $\si^A$ are a covariant basis (the intrinsic vielbein basis)
for the dual of the tangent space, called 
cotangent space (the space of 1-forms). The ``coordinate" 
cotangent basis dual to the $\pdyN$ vectors is given by the 
differentials $dy^M$ ($dy^M(\pdyN)=\de^M_N$). The components of 
$\si^A(y)$ on the coordinate basis are denoted $e_M^{~~A} (y)$:
\eq
\si^A (y)=e_M^{~~A} (y)~ dy^M   \label{2.9}
\en
\noi From the duality of the tangent and cotangent bases:
\eqa
& & e_M^{~~A}~e_B^{~~M}=\de_B^A  \\
& & e_A^{~~M}~e_N^{~~A}=\de_N^M \label{2.10}
\ena
\sk
\noi {\bf Note 1}: Substituting $t_A$ by
$e_A^{~~N} (y) \pdyN$  into the commutator (\ref{2.5}) leads to
the differential condition on $e_A^{~~M} (y)$:
\eq
 -2e_{[A}^{~~~N}~e_{B]}^{~~~M} \partial_N e_M^{~~C} = \CCAB
 \label{2.11}
 \en
{\bf Note 2}: computing the exterior derivative of $\si^A$, using eq.s 
(\ref{2.9}) and (\ref{2.11}) leads to the equations
\eq
d\si^A+{1 \over 2}{C^A}_{BC}\si^B \we \si^C = 0 \label{2.12}
\en
\noi These are called {\sl Cartan-Maurer equations}, and provide a dual 
formulation of Lie algebras in terms of the one-forms $\si^A$. It is 
immediate to verify that the closure of the exterior derivative ($d^
2=0$) is equivalent to the Jacobi identities for the structure 
constants:
\eq
C^A_{~~B[C} C^B_{~~DE]} =0  \label{2.13}
\en
\noi (apply $d$ to eq. (\ref{2.12})).
\sk
\noi {\bf Note 3}:
\sk
\noi Defining $\si (y) \equiv \si^A (y) T_A$ 
the Cartan-Maurer eq.s (\ref{2.12}) 
take the form
\eq
d\si+\si\we\si=0 \label{2.14}
\en
The Lie-valued one-form $\si(y)$ can also be constructed directly 
from the group element $y$:
\eq
\si(y)=y^{-1} dy  \label{2.15}
\en
It is easy to verify that (\ref{2.15}) satisfies the Cartan-Maurer equation (\ref{2.14}) (use $dy^{-1}=-y^{-1}dy~y^{-1}$). Moreover, it takes the same value 
as $e_M^{~~A} dy^M~T_A$ at the origin $y=0$. Indeed from the definition 
of $e_A^{~~M}$ in eq. (\ref{2.3}) one sees that $e_A^{~~M} (y=0)= \de_A^M$, 
and therefore $e_M^{~~A} (0) dy^M ~T_A=dy^A~T_A$. This value coincides 
with $y^{-1}dy|_{y=0}$ since $y^{-1}|_{y=0} =$[group unit], and 
$dy|_{y=0}=dy^A T_A$ (from (\ref{2.2})). This 
observation suffices to conclude that $y^{-1} dy$ is equal to 
$e_M^{~~A} (y) dy^M T_A$.
\sk
\noi {\bf Soft group manifold}
\sk
Consider a smooth deformation $\Gt$ of the group manifold $G$. Its 
vielbein field is given by the intrinsic cotangent basis, 
defined for any differentiable manifold:
\eq
\mu^A (y)=\mu_M^{~~A} (y) dy^M \label{3.1}
\en
\noi (In this Appendix we use the symbol $\mu$ for the 
``soft" vielbein). In general $\mu^A$ does not satisfy the Cartan-Maurer 
equations any more, so that
\eq
d\mu^A+{1 \over 2} \CABC \mu^B \we \mu^C \equiv R^A \not= 0 \label{3.2}
\en
\noi The extent of the deformation $G \rightarrow \Gt$ is measured by 
the curvature two-form $R^A$. $R^A = 0$ implies $\mu^A=\si^A$ and 
viceversa.
\sk
Applying the external derivative $d$ to the definition (\ref{3.2}), using $d^2
=0$ and the Jacobi identities on $\CABC$, yields the Bianchi identities
\eq
(\nabla R)^A \equiv dR^A - \CABC R^B \we \mu^C =0 \label{3.3}
\en
\sk
\noi {\bf Diffeomorphisms and Lie derivative}
\sk
First we discuss the variation under diffeomorphisms of the vielbein
field $\mu^A(y)$:
\eqa
     & & \mu^A (y+\de y)-\mu^A(y)=\de [\mu_M^{~~A}(y)dy^M]= \nonumber \\
     & & =(\part_N \mu_M^{~~A}) \de y^N dy^M+ \mu_M^{~~A} (\part_N \de y^M) 
        dy^N= \nonumber\\
     & &=dy^N[\part_N \de y^A+\de y^M(\part_M \mu_N^{~~A}-\part_N \mu_M
       ^{~~A})]= \nonumber\\
     & & = d \de y^A-2 \mu^B \de y^C (d\mu^A)_{BC}=d(\iota_{\de y}\mu^A)
       +\iota_{\de y} d \mu^A \label{3.7} 
       \ena
\noi where
\eq
\de y^A \equiv \de y^M \mu_M^{~~A},~~~\de y \equiv \de y^M 
\part_M,~~~d\mu^A\equiv (d\mu^A)_{BC} \mu^B \we \mu^C, \label{3.8}
\en
\noi and the contraction $\iota_{t}$ along a tangent vector ${t}$ 
is defined on p-forms
\eq
\om_{(p)}=\om_{B_1...B_p}\mu^{B_1} \we ... \we \mu^{B_p} 
\en
\noi as  
\eq
\iota_{t}~ \om_{(p)}=p~ t^A \om_{AB_2...B_p}\mu^{B_2}
 \we ... \we \mu^{B_p}
    \label{3.9}
    \en
\noi Note that $\iota_{t}$ maps p-forms into ($p-1$)-forms. 
The operator 
\eq
\ell_{t} \equiv d~ \iota_{t} + \iota_{t} ~d  \label{3.10}
\en
\noi is called the {\sl Lie derivative} along the tangent vector $t$ 
and maps p-forms into p-forms. 
As shown in eq. (3.7), the Lie derivative of the one-form 
$\mu^A$ along $\de y$ gives 
its variation under the diffeomorphism $y \rightarrow y+\de y$.   
This holds true for any p-form.
\sk
We now rewrite the variation $\de \mu^A$ of eq. (\ref{3.7}) in a suggestive 
way, by adding and subtracting $\CABC \mu^B  \de y^C$ :
\eqa
   \de \mu^A &= & d \de y^A + \CABC \mu^B  \de y^C - 2 \mu^B \de y^C 
(d\mu^A)_{BC} - \CABC \mu^B \de y^C \\
             &=& (\nabla \de y)^A + \iota_{\de y} R^A \\
      \label{3.11}
      \ena
\noi where we have used the definition (\ref{3.2}) for the curvature, and
the $G$-covariant derivative $\nabla$ acts on $\de y^A$ as
\eq
(\nabla \de y)^A \equiv d \mu^A + \CABC \mu^B \de y^A \label{3.12}
\en

\vfill\eject

{\bf The algebra of Lie derivatives}
\sk
The algebra of diffeomorphisms
 is given by the commutators of Lie derivatives:
 \eq
 \left[ \ell_{ \epsi^A_1 t_A},\ell_{\epsi_2^B t_B} \right]= \ell_{\epsi^C_3 t_C} \label{aldiffLie}
\en
with
\eq
\epsi_3^C=\epsi^A_1 \partial_A
\epsi^C_2 - \epsi^A_2 \partial_A \epsi^C_1 - 2 \epsi^A_1 \epsi^B_2
\Rcal^C_{AB}
\en
and
 \eq \Rcal^C_{AB} \equiv R^C_{AB}-\unmezzo C^C_{AB}
\label{Rcal}
\en
The components $R^A_{BC}$ are defined by $R^A = R^A_{BC} \mu^B \we \mu^C$. The closure of the algebra requires the Bianchi identities (\ref{3.3}), that we can rewrite in the form
\eq
\part_{[B} \Rcal^A_{CD]}+2~\Rcal^A_{E[B} \Rcal^E_{CD]} =0
\label{BianchiRcal}
\en
To prove (\ref{aldiffLie}) just apply both sides of the equation to the basic (soft) vielbein $\mu$.
\sk
\sk

 \sect{Spinors in $d=s+t$ dimensions}
  
   $\bullet~~~ \eta_{ab} = (\underbrace{1,1,\cdots,1}_t, \underbrace{-1,-1, \cdots -1}_s)$
\sk
\noi $\bullet~~~\{ \ga_a,\ga_b \} = 2 \eta_{ab}$
\sk
\noi $\bullet~~~$ $\ga_{a_1 \cdots a_n}$ = antisymmetrized product of $n$ gamma matrices, with weight 1. 
\sk
\noi $\bullet~~~$ A matrix representation of $\ga$'s can be made unitary by choice of basis
$\rightarrow$ 

``time" $\ga_a$ are hermitian, ``space" $\ga_a$ are antihermitian.
\sk
\noi $\bullet~~~$ Explicit representation: 
\eqa
& & \ga_1 = \si_1 \otimes \One ~\otimes \cdots  \cdots \cdots \otimes \One  \nonumber \\
& & \ga_2 = \si_2 \otimes \One ~ \otimes \cdots \cdots \cdots \otimes \One  \nonumber \\
& & \ga_3 = \si_3 \otimes \si_1 \otimes \cdots \cdots \cdots  \otimes \One  \nonumber \\
& & \ga_4 = \si_3 \otimes \si_2 \otimes \cdots  \cdots \cdots \otimes \One  \nonumber \\
& & ~~~~~~~~~~\vdots \nonumber \\
& & \ga_d = \si_3 \otimes \si_3 \otimes \cdots \otimes \si_3 \otimes \si_2,~~~d=2p  \nonumber \\
& & \ga_d = \si_3 \otimes \si_3 \otimes \cdots \otimes \si_3 \otimes \si_3,~~~d=2p+1  
\ena
~~~~~$\si_i$ = Pauli matrices, and multiply space $\ga$'s by $i$.
\sk
\noi $\bullet~~~$ the product of all gammas $\ga \equiv \ga_1 \ga_2 \cdots \ga_d$
 is proportional to the unit matrix when $d=2p+1$. When $d=2p$ , $\ga$ anticommutes with every $\ga_a$.
 For any $d$, $\ga^2=(-1)^{s+p}~\One$. 
 \sk
\noi $\bullet~~~$ In $d=2p$ there is only 1 irrep of the Clifford algebra\footnote{more precisely irrep
of the finite group $\Gamma (t,s)$ with elements $\pm I, \pm \ga_a, \pm \ga_{ab}  \cdots, \pm \ga_{a_1 \cdots a_d}$}, in $d=2p+1$ there are
two inequivalent irreps (if $\ga_a$ is in one irrep, the other inequivalent irrep is given by $-\ga_a$).
\sk
\noi $\bullet~~~$ $\ga_a$, $-\ga_a$, $\ga_a^T$ and $\ga_a^\dagger$ satisfy the same
Clifford algebra. Thus in $d=2p$ their irreps are all equivalent, while in $d=2p+1$ they
are equivalent up to a sign. Therefore in any $d$ we have
\eqa
& & \ga_a^T = \pm C_{\pm} \ga_a C^{-1}_{\pm} \label{defC}\\
& & \ga_a^\dagger = \pm A_{\pm} \ga_a A^{-1}_{\pm} \label{defA}
\ena
In the explicit representation, the solution for $C$ and $A$ matrices is unique 
(up to a factor) in $d=2p+1$ and twofold in $d=2p$:
\sk
{\bf $d$ odd:}
\eqa
& & C=\ga_1 \ga_3 \ga_5  \cdots \ga_{d} \\
& & A=\ga_1 \ga_2  \ga_3 \cdots \ga_{t}
\ena
~~~~~$C$ is either $C_+$ or $C_-$, depending on $s$ and $t$. The same holds for $A$.
\sk
{\bf $d$ even:}
\eqa
& & C_I~=\ga_1 \ga_3 \ga_5  \cdots \ga_{d-1} \\
& & C_{II}=\ga_2 \ga_4 \ga_6  \cdots \ga_{d} \\
& & A_{I}~=\ga_1 \ga_2  \ga_3 \cdots \ga_{t}\\
& & A_{II}=\ga_{t+1} \ga_{t+2}  \ga_{t+3} \cdots \ga_{d}
\ena
~~~~~If $C_I$ is $C_+$, then $C_{II}$ is $C_-$, and viceversa, depending on $s$ and $t$.

The same holds for $A$. Note that $A_I$ reproduces the usual  $\ga_1$ for $t=1$. In the following, we will always use $A=A_I$.
\sk
\noi $\bullet~~~$ Transposition properties of $\ga_a$ matrices can be deduced in the explicit representation
directly from Pauli matrices ($\si_1,\si_3$ symmetric, $\si_2$ antisymmetric), so that
$\ga_a$ is symmetric if $a$ is odd, antisymmetric for $a$ even. Consequently one has
\eq
C^T = \xi C \label{Ctransposed}
\en
For $d$ {\bf odd} one finds
\eq
\xi = (-1)^{[(d+1)/4]} 
\en
where $[\cdots ]$ denotes the integer part, and for $d$ {\bf even}:
\eq
\xi_I=(-1)^{[(d+1)/4]} ,~~~~~~\xi_{II}=(-1)^{[(d+2)/4]}  
\en

\sk
\noi $\bullet~~~$ defining
\eq
Q_{I,II} \equiv (A^{-1})^T C_{I,II}
\en
we find for $\ga^*_a$ a relation analogous to (\ref{defC}),(\ref{defA}):
\eq
\ga^*_a= \chi Q \ga_a Q^{-1} \label{gammastar}
\en
with
\eq
\chi_I=(-1)^{[(s-t+1)/2]},~~~\chi_{II}=(-1)^{[(t-s+1)/2]}
\en
and
\eq
Q_IQ^*_I=(-1)^{[(s-t+1)/4]},~~~Q_{II}Q^*_{II}=(-1)^{[(t-s+1)/4]}  \label{QstarQ}
\en
\sk
\noi $\bullet~~~$ Applying a Lorentz transformation $\La_a^{~b} \in SO(t,s)$ on the vector index of $\ga_b$
yields
\eq
(\La \ga)_a = \La_a^{~b} \ga_b
\en
and since  $(\La \ga_a)$ is still a representation of the
Clifford algebra we must have:
\eq
(\La \ga)_a = S^{-1} (\La)  \ga_a S (\La)
\en
Taking $\La_a^{~b}$ infinitesimal, i.e. $\La_a^{~b} = \de^b_a + \epsi_a^{~b}$ we find
\eq
S(\La) = \One  +{1 \over 4} \epsi^{ab} \ga_{ab}
\en
by using
\eq
[\ga_{ab},\ga_c]= 2 \eta_{bc} \ga_a - 2\eta_{ac} \ga_b
\en
valid in any dimension. The set of matrices $S(\La)$ forms the Spin group $Spin(t,s)$, and
\eq
SO(t,s)={Spin(t,s) \over Z_2}
\en
The following relations are easy to prove:
\eq
A=SAS^\dagger,~~~C=S^T C S,~~~\ga S = S \ga,~~~Q_{I,II}= S^* Q_{I,II} S^{-1}
\en
\noi $\bullet~~~$ By definition a spinor transforms under a Lorentz transformation $\La$ as 
\eq
\psi'= S(\La) \psi \label{Lorentzonspinor}
\en
\noi $\bullet~~~$The $S(\La)$ are not unitary in general. The matrix representation $S(\La)$ 
is reducible in $d=2p$ since all $S(\La)$ commute with $\ga$: there are two distinct irreps for
spinors, with dimension $2^{p-1}$. On the other hand, for $d=2p+1$ the spinor representation
is irreducible. Both irreps of $\Ga(t,s)$, connected by $\ga_a \rightarrow - \ga_a$, lead to the same
spinor irrep of dimension $2^p$ since $\ga_{ab}$ is not changed by $\ga_a \rightarrow - \ga_a$.
\sk
\noi $\bullet~~~$ The {\bf Dirac conjugate}:
\eq
\psibar \equiv \psi^\dagger A
\en
transforms as
\eq
\psibar ' = \psibar S^{-1} (\La)
\en
\noi $\bullet~~~$The currents
\eq
j_{a_1 \cdots a_n} = \psibar \ga_{a_1 \cdots a_n} \psi
\en
transform under (\ref{Lorentzonspinor}) as tensors:
\eq
j'_{a_1 \cdots a_n} = \La_{a_1}^{~b_1} \cdots  \La_{a_n}^{~b_n} j_{b_1 \cdots b_n} 
\en
\noi $\bullet~~~$  The {\bf charge conjugated} spinor} is defined by
\eq
\psi^c \equiv C \psibar^T
\en
If  $\psi$ satisfies the Dirac equation
\eq
(i \ga^a \part_a - e \ga^a A_a -m) \psi =0
\en
then $\psi^c$ satisfies
\eq
(i \ga^a \part_a + e \ga^a A_a - \chi m) \psi ^c=0
\en
with a change of sign of 
the electric charge, and $\chi$ as defined in (\ref{gammastar}). For this reason
$C$ is also called the {\it charge conjugation} matrix.
\sk
\noi $\bullet~~~$A {\bf Majorana spinor} is defined to satisfy:
\eq
\psi^\dagger A = \psi^T C
\en
or equivalently
\eqa
& & \psi^* = \xi Q \psi,  \label{psistar}\\
& & \psi^c = \alpha \psi
\ena
with $\xi$ given in (\ref{Ctransposed}) and $\alpha = \pm 1$ defined by $(C^{-1})^T = \alpha C$.
Iterating (\ref{psistar}) one finds the condition on $Q$:
\eq
QQ^*=\One
\en
implying 
\eqa
& & [(s-t+1)/4] = 0 ~(mod~2) {\rm ~~~~for ~Q_I ~Majorana ~spinors} \\
& & [(t-s+1)/4] = 0 ~(mod~2) {\rm ~~~~for ~Q_{II}  ~Majorana~ spinors}
\ena
cf.  (\ref{QstarQ}). Therefore, defining
\eq
f \equiv t-s 	\label{fdef}
\en
 one has $Q_I$ Majorana spinors for $f=-2,-1,0,1$ (mod 8) and 
 $Q_{II}$ Majorana spinors for $f=-1,0,1,2$ (mod 8).
 \sk
 \noi $\bullet~~~$ {\bf Self-dual tensors}: 
 \eq
 {F}_{a_1 \cdots a_p} = {1 \over p!} \epsi_{a_1 \cdots a_p b_1 \cdots b_p} F^{b_1 \cdots b_p} \label{selfdual}
 \en
Iterating (\ref{selfdual}) and using
\eq
\epsi_{a_1 \cdots a_r c_1 \cdots c_q} \epsi^{b_1 \cdots b_r c_1 \cdots c_q} = (-1)^s ~r! ~q! ~\de^{b_1 \cdots b_r}_{a_1 \cdots a_r}
\en
implies
\eq
 {F}_{a_1 \cdots a_p} = (-1)^{p+s} ~ {F}_{a_1 \cdots a_p} = (-1)^{f/2}~ F_{a_1 \cdots a_p}
 \en
 Therefore selfdual (or antiselfdual) tensors exist only if $f=0$ mod 4.
\sk
\noi $\bullet~~~$ {\bf Weyl spinors}. Defining a $d$-dimensional analog of $\ga_5$
\eq
\Ga = (-1)^{f/4} \ga,~~~\Ga^2=\One
\en
Weyl or anti-Weyl spinors are defined in any even dimension by:
\eq
\psi = \pm \Ga \psi
\en
Spinors satisfying both the Majorana and the Weyl condition exist in even
dimensions only if $\psi^c$ has the same chirality as $\psi$, since $\psi^c=\al \psi$ for Majorana
spinors. Using the explicit representation one can prove that 
if $\psi$ has chirality +1 (-1) , then $\psi^c$ has chirality $(-1)^{f/2}$ ( $-(-1)^{f/2})$. 
Therefore $f=0$ mod 4 is a necessary condition for Weyl spinors to be Majorana. 
Combining this condition with the conditions for the existence of $Q_I$ or $Q_{II}$ Majorana spinors
given after eq. (\ref{fdef}), one finds that MW spinors exist if and only if $f=0$ mod 8.
\sk
\noi $\bullet~~~$ Transposition properties of the matrices $C \ga_{(n)}$, where
$\ga_{(n)}$ is a shortand notation for $\ga_{a_1 \cdots a_n}$, are important to know,
since they determine which currents 
\eq
\psibar \ga_{(n)} \psi = \psi^\alpha C_{\alpha\ga} \ga^{~~\ga}_{(n)~\beta} \psi^\beta
\en
 can exist for Majorana spinors $\psi$. If $\psi$ is a zero-form (one-form), the current 
 $\psibar \ga_{(n)} \psi$ exists if the matrix  $C \ga_{(n)}$
 is antisymmetric (symmetric), since $\psi^\al \psi^\be$ is antisymmetric (symmetric) in
 $\al,\be$. In general
 \eq
 (C \ga_{(n)})^T =  \eta^n (-1)^{n(n-1)/2} \xi ~C \ga_{(n)}
 \en
 where $\xi$ is given after (\ref{Ctransposed}) and $\eta$ is $+1$ for $C_+$ and $-1$ for $C_-$.
 \sk
 
 A table for Minkowski signature ($t=1,s=d-1$) follows. In computing $ (C \ga_{(n)})^T$ we have chosen $C_I$ for $d=2,4,10,12$ and $C_{II}$ for $d=6,8$. The table also lists the properties of $C_I,C_{II}$, and the types of Majorana spinors ($Q_I$ and/or $Q_{II}$) in $2 \le d \le 12$.
 \sk  
 
 \vfill\eject
 
 \centerline{Table 2: properties of gamma matrices and spinors in $d$ dimensions}
 \centerline {with Minkowski signature ($t=1$, $s=d-1$)}
 \sk
 \begin{tabular}{|c|c|c|c|c|c|} \hline $d$ & Symm &Antisymm& $C_I$& $C_{II}$ &Majorana  \\   & $C \ga_{(n)}$ & $C \ga_{(n)}$ & & & \\   &for $n=$ & for $n=$ & & &  \\\hline 2 & 0,1 & 2 & $C_+^T=C_+$  & $C_-^T= -C_-$ & $Q_I,Q_{II}$  \\\hline 3 & 1,2 & 0 & $C_-^T=-C_-$ & & $Q_I$  \\\hline 4 & $1,2$& $0,3,4 $& $C_-^T=-C_-$ & $C_+^T=-C_+$ & $Q_I$   \\\hline 5 & 2 & 0,1 & $C_+^T=-C_+$ &  &  \\\hline 6 & 0,3,4 & 1,2,5,6 & $C_+^T=-C_+$ & $C_-^T= C_-$ & \\\hline 7 & 0,3 & 1,2 & $C_-^T= C_-$  &  & \\\hline 8 & 0,1,4,5,8 & 2,3,6,7 & $C_-^T= C_-$ & $C_+^T= C_+$ & $Q_{II}$  \\\hline 9 & 0,1,4,5 & 2,3 &  $C_+^T= C_+$ &  & $Q_I$ \\\hline 10 & 1,2,5,6,9,10 & 0,3,4,7,8 &  $C_+^T= C_+$ &  $C_-^T= -C_-$ & $Q_I,Q_{II}$  \\\hline 11 & 1,2,5 & 0,3,4 & $C_-^T= C_-$ &  & $Q_I$\\\hline 12 & 1,2,5,6,9,10 & 0,3,4,7,8,11,12 & $C_-^T= -C_-$ & $C_+^T= -C_+$ 
 & $Q_I$ \\\hline \end{tabular}

\sk
A nice summary of the properties of spinors in $d=t+s$ is given by the ``spinor clock" 
designed by Tullio Regge in ref. \cite{gm14}, reproduced in the following Figure:

\includepdf[fitpaper=true]{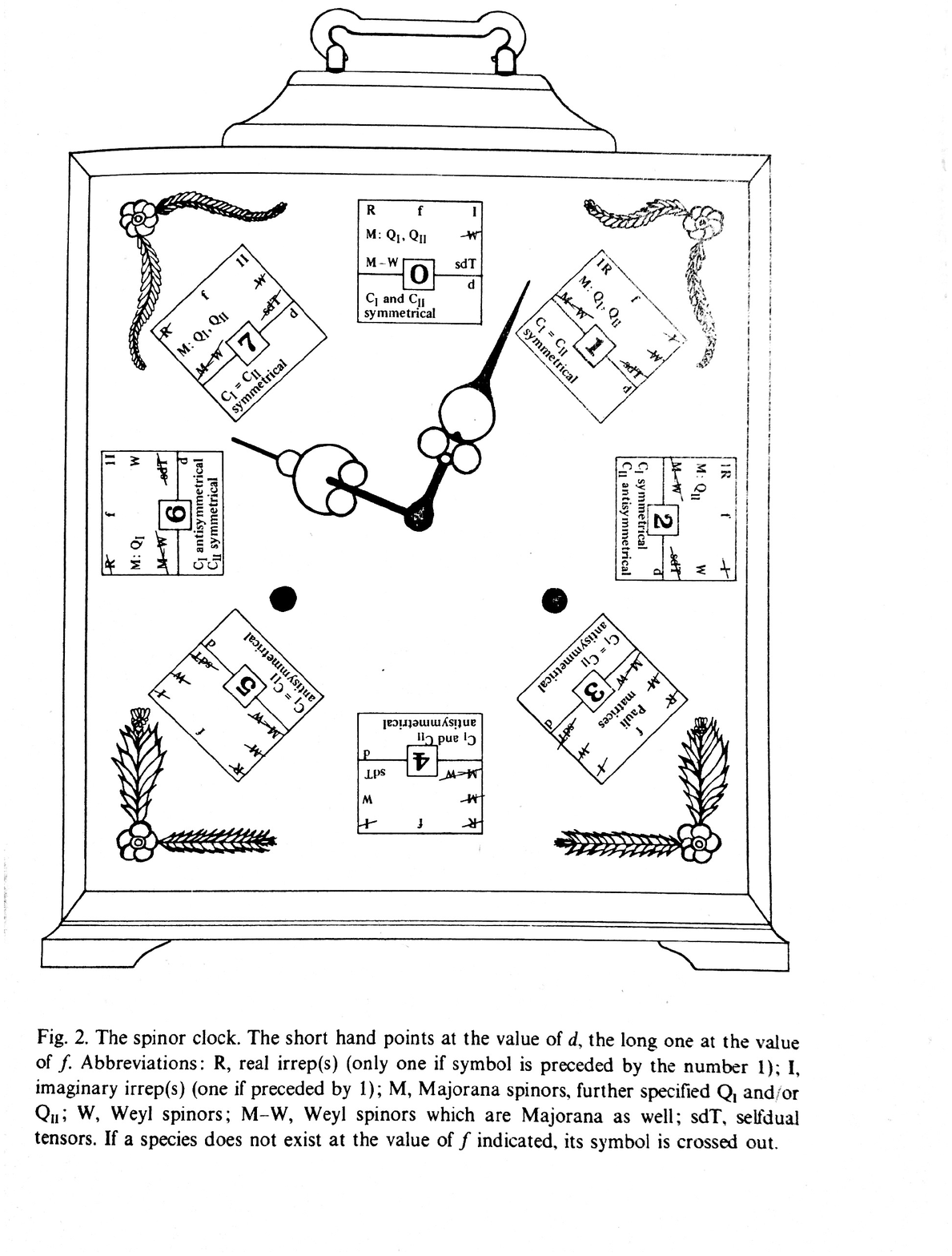}

 \sect{$\gamma$ matrices in $d=2+1$}

We adopt the traditional numbering $0,1,2$ instead of $1,2,3$.
\eq
 \ga_0 =
 \left(
\begin{array}{cc}
  0&   1  \\
  1&     0
\end{array}
\right)
,~~~\ga_1=
\left(
\begin{array}{cc}
  0&   1  \\
  -1&     0
\end{array}
\right)
,~~~\ga_2=
\left(
\begin{array}{cc}
  i &  0    \\
  0&  -i
\end{array}
\right)\en
\eqa
& & \eta_{ab} =(1,-1,-1),~~~\{\ga_a,\ga_b\}=2 \eta_{ab},~~~[\ga_a,\ga_b]=2 \ga_{ab}= -2 i \epsi_{abc} \ga^c, \\
& & \epsi_{012} =  \epsi^{012}=1, \\
& & \ga_a^\dagger = \ga_0 \ga_a \ga_0,~~\ga_a^T= - C \ga_a C^{-1}
\ena
\eq
C= i \ga_0 \ga_2 =  \left(
\begin{array}{cc}
  0&   1 \\
  -1&     0
\end{array}
\right) ~~~\longrightarrow ~~~C_{\al\be}=\epsi_{\al\be}
\en

\subsection{Useful identities}

\eqa
 & &\ga_a\ga_b= \ga_{ab}+\eta_{ab}= - i \epsi_{abc} \ga^c + \eta_{ab}\\
 & &\ga_{ab} \ga_c=\eta_{bc} \ga_a - \eta_{ac} \ga_b - i\epsi_{abc}\\
 & &\ga_c \ga_{ab} = \eta_{ac} \ga_b - \eta_{bc} \ga_a -i \epsi_{abc}\\
 & &\ga_a\ga_b\ga_c= \eta_{ab}\ga_c + \eta_{bc} \ga_a - \eta_{ac} \ga_b - i \epsi_{abc}\\
 & &\ga^{ab} \ga_{cd} = - 4 \de^{[a}_{[c} \ga^{b]}_{~~d]} - 2 \de^{ab}_{cd}
  \ena
 \noi where $\de^{ab}_{cd}
 = \unmezzo (\de^a_c \de^b_d - \de^a_d \de^b_c)$, and index antisymmetrizations in square brackets have weight 1.

\subsection{Fierz identity for two Majorana one-forms}

\eq
\psi \psibar = {1 \over 2} (\psibar \ga^a \psi ) \ga_a
\en
As a consequence 
\eq
\ga_a \psi \psibar \ga^a \psi =0   \label{Fierz3d}
\en

 \sect{$\gamma$ matrices in $d=3+1$}
 
 We use the traditional numbering $0,1,2,3$ instead of $1,2,3,4$.
\eqa
& & \eta_{ab} =(1,-1,-1,-1),~~~\{\ga_a,\ga_b\}=2 \eta_{ab},~~~[\ga_a,\ga_b]=2 \ga_{ab}, \\
& & \ga_5 \equiv  -i \ga_0\ga_1\ga_2\ga_3,~~~\ga_5 \ga_5 = 1,~~~\epsi_{0123} = - \epsi^{0123}=1, \\
& & \ga_a^\dagger = \ga_0 \ga_a \ga_0, ~~~\ga_5^\dagger = \ga_5 \\
& & \ga_a^T = - C \ga_a C^{-1},~~~\ga_5^T = C \ga_5 C^{-1}, ~~~C^2 =-1,~~~C^T =-C
\ena

\subsection{Useful identities}
\eqa
 & &\ga_a\ga_b= \ga_{ab}+\eta_{ab}\\
 & & \ga_{ab} \ga_5 = - {i \over 2} \epsilon_{abcd} \ga^{cd}\\
 & &\ga_{ab} \ga_c=\eta_{bc} \ga_a - \eta_{ac} \ga_b +i \epsi_{abcd}\ga_5 \ga^d\\
 & &\ga_c \ga_{ab} = \eta_{ac} \ga_b - \eta_{bc} \ga_a +i \epsi_{abcd}\ga_5 \ga^d\\
 & &\ga_a\ga_b\ga_c= \eta_{ab}\ga_c + \eta_{bc} \ga_a - \eta_{ac} \ga_b +i \epsi_{abcd}\ga_5 \ga^d\\
 & &\ga^{ab} \ga_{cd} = i \epsi^{ab}_{~~cd}\ga_5 - 4 \de^{[a}_{[c} \ga^{b]}_{~~d]} - 2 \de^{ab}_{cd}
 \ena

 \subsection{Charge conjugation and Majorana condition}

\eqa
 & &   {\rm Dirac~ conjugate~~} \psibar \equiv \psi^\dagger
 \ga_0\\
 & &  {\rm Charge~ conjugate~spinor~~} \psi^c = C (\psibar)^T  \\
 & & {\rm Majorana~ spinor~~} \psi^c = \psi~~\Rightarrow \psibar =
 \psi^T C
 \ena

\subsection{Fierz identity for two spinor one-forms}
\eq
 \psi  \chibar = \unquarto [ (\chibar  \psi) 1 + (\chibar \ga_5  \psi) \ga_5 + (\chibar \ga^a  \psi) \ga_a + (\chibar \ga^a \ga_5  \psi) \ga_a \ga_5  - \unmezzo (\chibar \ga^{ab}  \psi) \ga_{ab}]
 \en
 \subsection{Fierz identity for two Majorana spinor one-forms}
 \eq
 \psi  \psibar = \unquarto [  (\psibar \ga^a  \psi) \ga_a  - \unmezzo (\psibar \ga^{ab}  \psi) \ga_{ab}]
 \en
 \noi As a consequence
 \eq
\ga_a \psi \psibar \ga^a \psi =0,~~~ \psi \psibar \ga^a \psi- \ga_b \psi \psibar \ga^{ab} \psi=0 \label{Fierz4d}
\en

  \sect{$\gamma$ matrices in $d=4+1$}
  
  \eqa
& & \eta_{ab} =(1,-1,-1,-1,-1),~~~\{\ga_a,\ga_b\}=2 \eta_{ab},~~~[\ga_a,\ga_b]=2 \ga_{ab}, \\
& & \ga_0\ga_1\ga_2\ga_3\ga_4=-1,~~~\epsi_{01234} =  \epsi^{01234}=1, \\
& & \ga_a^\dagger = \ga_0 \ga_a \ga_0,  \\
& & \ga_a^T =  C \ga_a C^{-1}, ~~~C^2 =-1,~~~C^\dagger=C^T =-C
\ena

\subsection{Useful identities}

\eqa
 & &\ga_a\ga_b= \ga_{ab}+\eta_{ab}\\
 & & \ga_{abc}  = {1 \over 2} \epsilon_{abcde} \ga^{de}\\
 & & \ga_{abcd}  = - \epsilon_{abcde} \ga^{e}\\
 & &\ga_{ab} \ga_c=\eta_{bc} \ga_a - \eta_{ac} \ga_b +{1 \over 2} \epsilon_{abcde} \ga^{de}\\
 & &\ga_c \ga_{ab} = \eta_{ac} \ga_b - \eta_{bc} \ga_a+{1 \over 2} \epsilon_{abcde} \ga^{de}\\
 & &\ga^{ab} \ga_{cd} = - \epsi^{ab}_{~~cde}\ga^e - 4 \de^{[a}_{[c} \ga^{b]}_{~~d]} - 2 \de^{ab}_{cd}
 \ena
\noi where
$\delta^{ab}_{cd} \equiv \frac{1}{2}(\delta^a_c\delta^b_d-\delta^b_c\delta^a_d)$, $\delta^{rse}_{abc} \equiv  {1 \over 3!} (\de^r_a \de^s_b \de^e_c$ + 5 terms), 
and index antisymmetrization in square brackets has total weight $1$.

\vfill\eject
\end{document}